\documentclass[twocolumn,twoside,showkeys,showpacs,preprintnumbers,superscriptaddress,amsmath,amssymb,prc]{revtex4-2}

\usepackage{graphicx}% Include figure files
\usepackage{dcolumn}% Align table columns on decimal point
\usepackage{bm}% bold math
\usepackage[version=4]{mhchem}
\usepackage{subcaption}
\usepackage[colorlinks=true,linkcolor=black,citecolor=blue,urlcolor=blue,]{hyperref}
\usepackage{appendix}
\usepackage{ulem} 
\usepackage[justification=raggedright,singlelinecheck=false]{caption}
\usepackage{booktabs}

\usepackage{pstricks}
\usepackage{amsmath, amsthm, amssymb}
\usepackage{changes}

\def\snsn{ $\rm ^{124}Sn$+$\rm ^{124}Sn$} 
\def\snnucl {$\rm ^{124}Sn$} 

\def\ssdtwo {\texttt{SSD M2}}

\begin{document}

% \title{Probing Nucleon-Nucleon Short Range Correlations by Bremsstrahlung $\gamma$-rays in Heavy Ion Collisions} 

\title{Experimental Study of Bremsstrahlung Gamma Ray Emission  and Short-Range Correlations in $^{124}$Sn+$^{124}$Sn Collisions at 25 MeV/u}

\author{Junhuai Xu}
\email{xjh22@mails.tsinghua.edu.cn}
\affiliation{Department of Physics, Tsinghua University, Beijing 100084, China}%

\author{Qinglin Niu}
\email{602023220053@smail.nju.edu.cn}
\affiliation {School of Physics, Nanjing University, Nanjing 210093, China}

\author{Yuhao Qin}%
\affiliation{Department of Physics, Tsinghua University, Beijing 100084, China}%

\author{Dawei Si}
\affiliation{Department of Physics, Tsinghua University, Beijing 100084, China}%

\author{Yijie Wang}
\affiliation{Department of Physics, Tsinghua University, Beijing 100084, China}%
%\affiliation{Department of Physics, Tsinghua University, Beijing 100084, China}%

\author{Sheng Xiao}
\affiliation{Department of Physics, Tsinghua University, Beijing 100084, China}%

\author{Baiting Tian}
\affiliation{Department of Physics, Tsinghua University, Beijing 100084, China}%

\author{Zhi Qin}
\affiliation{Department of Physics, Tsinghua University, Beijing 100084, China}%

\author{Haojie Zhang}
\affiliation{Department of Physics, Tsinghua University, Beijing 100084, China}%

\author{Boyuan Zhang}
\affiliation{Department of Physics, Tsinghua University, Beijing 100084, China}%

\author{Dong Guo}
\affiliation{Department of Physics, Tsinghua University, Beijing 100084, China}%

\author{Minxue Fu}
\affiliation{Department of Physics, Tsinghua University, Beijing 100084, China}%

\author{Xiaobao Wei}
\affiliation {Institute of Particle and Nuclear Physics, Henan Normal University, Xinxiang 453007, China}

\author{Yibo Hao}
\affiliation {Institute of Particle and Nuclear Physics, Henan Normal University, Xinxiang 453007, China}

\author{Zengxiang Wang}
\affiliation {Institute of Particle and Nuclear Physics, Henan Normal University, Xinxiang 453007, China}

\author{Tianren Zhuo}
\affiliation {Institute of Particle and Nuclear Physics, Henan Normal University, Xinxiang 453007, China}

\author{Chunwang Ma}
\affiliation {Institute of Particle and Nuclear Physics, Henan Normal University, Xinxiang 453007, China}
\affiliation {Institute of Nuclear Science and Technology, Henan Academy of Sciences, Zhengzhou, 450015, China}

\author{Yuansheng Yang}
\affiliation {Institute of Modern Physics, Chinese Academy of Sciences, Lanzhou 730000, China}

\author{Xianglun Wei}
\affiliation {Institute of Modern Physics, Chinese Academy of Sciences, Lanzhou 730000, China}

\author{Herun Yang}
\affiliation {Institute of Modern Physics, Chinese Academy of Sciences, Lanzhou 730000, China}

\author{Peng Ma}
\affiliation {Institute of Modern Physics, Chinese Academy of Sciences, Lanzhou 730000, China}

\author{Limin Duan}
\affiliation {Institute of Modern Physics, Chinese Academy of Sciences, Lanzhou 730000, China}

\author{Fangfang Duan}
\affiliation {Institute of Modern Physics, Chinese Academy of Sciences, Lanzhou 730000, China}

\author{Kang Wang}
\affiliation {Institute of Modern Physics, Chinese Academy of Sciences, Lanzhou 730000, China}

\author{Junbing Ma}
\affiliation {Institute of Modern Physics, Chinese Academy of Sciences, Lanzhou 730000, China}

\author{Shiwei Xu}
\affiliation {Institute of Modern Physics, Chinese Academy of Sciences, Lanzhou 730000, China}

\author{Zhen Bai}
\affiliation {Institute of Modern Physics, Chinese Academy of Sciences, Lanzhou 730000, China}

\author{Guo Yang}
\affiliation {Institute of Modern Physics, Chinese Academy of Sciences, Lanzhou 730000, China}

\author{Yanyun Yang}
\affiliation {Institute of Modern Physics, Chinese Academy of Sciences, Lanzhou 730000, China}

\author{Mengke Xu}%
\affiliation {Shanghai Advanced Research Institute, Chinese Academy of Sciences, Shanghai 201210, China}
\author{Kaijie Chen}%
\affiliation {Shanghai Advanced Research Institute, Chinese Academy of Sciences, Shanghai 201210, China}
\author{Zirui Hao}%
\affiliation {Shanghai Advanced Research Institute, Chinese Academy of Sciences, Shanghai 201210, China}
\author{Gongtao Fan}%
\affiliation {Shanghai Advanced Research Institute, Chinese Academy of Sciences, Shanghai 201210, China}
\author{Hongwei Wang}%
\affiliation {Shanghai Advanced Research Institute, Chinese Academy of Sciences, Shanghai 201210, China}

\author{Chang Xu}
\email{cxu@nju.edu.cn}
\affiliation {School of Physics, Nanjing University, Nanjing 210093, China}

\author{Zhigang Xiao}
\email{xiaozg@tsinghua.edu.cn}
\affiliation{Department of Physics, Tsinghua University, Beijing 100084, China}%
\affiliation{Center for High Energy Physics, Tsinghua University, Beijing 100084, China}

\date{\today}

\begin{abstract}

Short-range correlation (SRC) in nuclei refers to nucleons forming temporally correlated pairs in close proximity, giving rise to the high momentum of the nucleons beyond the Fermi surface.  It has been reported that bremsstrahlung $\gamma$ production from neutron-proton process in heavy-ion reactions provides a potential probe to the SRC abundance in nuclei. In this paper, we present in detail the precision measurement of bremsstrahlung $\gamma$-rays in $\rm ^{124}Sn$+$\rm ^{124}Sn$ reactions at 25 MeV/u using the Compact Spectrometer for Heavy IoN Experiment (CSHINE). A comprehensive experimental and analysis framework is established to ensure the reliability and robustness of the extracted results. Background contributions are evaluated and subtracted using independent methods, and the consistency of the analysis is systematically validated. By comparing the experimental $\gamma$ spectrum with the Isospin-dependent Boltzmann-Uehling-Uhlenbeck simulations, the high momentum tail (HMT) fraction of $R_{\rm HMT}=(20 \pm 3)\%$ is derived in $^{124}$Sn nuclei. This work provides a detailed and validated experimental framework for extracting SRC information from bremsstrahlung $\gamma$-ray emission and demonstrates the feasibility of studying nucleon SRCs with high precision in low-energy heavy-ion collisions.

\end{abstract}

\maketitle

\section{Introduction}

Short-range correlation (SRC) \cite{Frankfurt:1988nt,CiofiDegliAtti:1989eg,Subedi:2008zz,Arrington:2011xs} of nucleons is a temporal fluctuation in which  nucleons form temporally correlated pairs in close proximity in  atomic nuclei. SRC plays an essential role in understanding the nuclear structure as well as to shed insight to the properties of dense nucleonic matter. Most traditional nuclear models are based on mean-field theory, in which nucleons are confined below the Fermi momentum and quark-level interactions are largely neglected. However, the EMC effect \cite{EuropeanMuon:1983wih} indicates that deviations in nucleon momentum distributions from shell model predictions stem from the underlying quark structure of nucleons \cite{Geesaman:1995yd,Kelly:1996hd,Norton:2003cb,Dickhoff:2004xx,Malace:2014uea}. This is exemplified by the fact that nuclear parton distribution functions (PDFs) in heavy nuclei deviate markedly from the naive sum of those in free nucleons \cite{Kovarik:2015cma,Eskola:2021nhw,Helenius:2021tof,AbdulKhalek:2022fyi,Segarra:2020gtj,Ruiz:2023ozv}, underscoring the importance of understanding the momentum distributions of quarks and gluons within bound nucleons and nuclei. SRCs, a ubiquitous feature of all nuclei, originate from fluctuations in the nuclear ground state and predominantly involve neutron-proton ($np$) pairs \cite{Piasetzky:2006ai,Subedi:2008zz,LabHallA:2014wqo,Hen:2014nza,CLAS:2018xvc,CLAS:2020rue,West:2020tyo}, driven by the strong tensor component of the nucleon-nucleon interaction at sub-Fermi-range distances \cite{Schiavilla:2006xx,Wiringa:2008dn,Wiringa:2013ala}. Notably, $np$ pairs often display significant spatial overlap, which may favor isosinglet, spin-singlet $ud$ diquark configurations at the quark level \cite{West:2020tyo}. These pairs also exhibit large relative momenta, contributing to the emergence of a high-momentum tail (HMT) in the nucleon momentum distribution beyond the Fermi surface \cite{Hen:2014nza,CLAS:2018yvt}. Since SRCs reflect dynamics beyond the reach of mean-field approximations, incorporating both short-range nuclear forces and potential quark-level effects \cite{nCTEQ:2023cpo}, they have become a focal point of contemporary theoretical and experimental research in nuclear physics.

Electron scattering experiments have proven to be particularly powerful in probing SRCs, owing to the well-understood nature of electromagnetic interactions. By carefully selecting kinematic conditions, complex final-state effects can be minimized \cite{Hen:2016kwk}, thereby improving the sensitivity to SRCs. In exclusive measurements \cite{JeffersonLabHallA:2007lly,Subedi:2008zz,LabHallA:2014wqo,CLAS:2010yvl,Hen:2014nza}, both the scattered electron and the two nucleons from the correlated pair are detected. A high-energy electron with large momentum transfer is used to knock out one nucleon from the SRC pair. Strong back-to-back correlations \cite{Frankfurt:1988nt} are observed in knocked-out nucleon pairs with momenta above the Fermi momentum, whereas no angular correlation is seen below it. In inclusive quasielastic (QE) electron scattering \cite{Frankfurt:1993sp,Arrington:1998ps,CLAS:2003eih,CLAS:2005ola,Fomin:2011ng,CLAS:2019vsb}, only the scattering electrons are measured and the momentum distribution of nucleons inside the nucleus can be probed. In particular, the scattering experiments related to low-energy transfer side of the QE peak can provide evidence for nucleons with high momenta \cite{Fomin:2011ng,Hen:2012fm}, indicating the presence of short-range correlated pairs.

Hadronic reactions also serve as one of the primary approaches for experimental studies of SRCs. In exclusive scattering experiments involving large momentum transfer proton-nucleus collisions \cite{Tang:2002ww,Piasetzky:2006ai}, the momentum of the struck proton can be reconstructed while the momentum of the correlated neutron is measured simultaneously. Under high momentum transfer conditions, the struck fast-bound proton can be described using the instantaneous approximation, which significantly enhances the resolution of the nuclear structure. Moreover, high-energy inverse kinematics scattering \cite{Patsyuk:2021fju}, where an incoming ion beam collides with a proton target, has emerged as a novel technique to investigate SRCs. By detecting the knocked-out protons and residual nuclear fragments, this method effectively suppresses complications from initial- and final-state interactions between hadrons \cite{Wakasa:2017rsk,CiofidegliAtti:2015lcu}, thereby improving the reliability of reconstructed particle distributions.

Heavy-ion collisions (HICs) provide the only means of producing and studying extreme states of nuclear matter in terrestrial laboratories. In the 1980s, high-energy photon emission observed in HICs emerged as a puzzling phenomenon \cite{Noll:1984fd,Njock:1986uzx,Stevenson:1986zz}, drawing growing interest due to its unclear origin. Subsequent experimental measurements have shown that a significant fraction of these $\gamma$-rays originate from bremsstrahlung processes induced by $np$ collisions in the early stage of HICs \cite{Bauer:1986zz,PhysRevC.53.R553,Remington:1987zza,Ko:1985hq,Jetter:1994bc,Wang:2020bzn,vanGoethem:2001hy,Maydanyuk:2013aqa}. Recent advances in the study of SRCs have substantially deepened our understanding of nuclear structure. Theoretical studies indicate that high-energy bremsstrahlung $\gamma$-ray production in HICs \cite{Xue:2016udl} is sensitive to the high-momentum components of nucleon momentum distributions generated by SRCs \cite{Xu:2012hf}, rendering hard photons in the $\gamma$-ray spectrum a promising probe of these correlations \cite{Xue:2016udl,Guo:2021zcs}. Notably, compared to the hadron probes proposed \cite{ Wei:2019jva,Huang:2025uvc,Hagel.WPCF2023}, photons interact only weakly with the nuclear medium and are largely unaffected by final-state interactions, making them a particularly clean and sensitive probe of short-range nuclear dynamics. Using the isospin-dependent Boltzmann-Uehling-Uhlenbeck (IBUU) transport model \cite{Li:2018lpy} with momentum-dependent interactions (MDI) \cite{Das:2002fr}, one can simulate $\gamma$-ray spectra from HICs by varying the fraction of high-momentum nucleons in the initial nuclear state. Comparing these simulations with experimental bremsstrahlung data enables the extraction of information on  SRC fractions in nuclei.

To explore the underlying dynamical mechanisms in intermediate-energy HICs, the Compact Spectrometer for Heavy Ion Experiment (CSHINE) \cite{Guan:2021tbi,Wang:2021jgu,Wei:2025lbj} has been constructed, maintained, and continuously upgraded at the Radioactive Ion Beam Line in Lanzhou (RIBLL-1). Through multiple beam campaigns, CSHINE has developed advanced detection technologies and delivered a wide range of physics results in key areas including the equation of states of nuclear matter, isospin evolutions \cite{Zhang:2017xtk,Wang:2021mrv,Wang:2022ysq}, short-range correlations in nuclei \cite{Xu:2025mvv,Xu:2024oct,Qin:2023qcn}, neutron-neutron interactions \cite{Si:2025eou}, and the interplay between fission dynamics and isospin dynamics \cite{Wang:2023kyp}. To date, CSHINE has completed five beam campaigns. As part of the detection system, a CsI(Tl)-based total absorption $\gamma$-ray spectrometer array (CSHINE-Gamma) \cite{Qin:2022mzp} was first deployed during the fourth experiment in 2022, which involved 25 MeV/u $^{86}$Kr + $^{124}$Sn reactions. In this experiment, CSHINE-Gamma successfully detected and analyzed $\gamma$-ray spectra emitted from HICs. The data revealed a non-zero HMT fraction, providing experimental indication of the presence of SRC-induced HMT components in nuclei at about 90\% confidence level \cite{Qin:2023qcn,Xu:2024oct}. However, this early study faced limitations due to the restricted energy coverage of CSHINE-Gamma and a narrow event reconstruction region confined to the central scintillator. These constraints hindered the ability to quantify the HMT fraction with high statistical confidence. Moreover, due to the use of an asymmetric reaction system, it was not possible to determine the SRC fraction for a specific nucleus.

In 2024, the fifth CSHINE experiment, involving the 25 MeV/u \snsn~ reactions, introduced significant upgrades to CSHINE-Gamma specifically improving the high-energy $\gamma$-ray detection. These improvements included an extended energy range, expanded event reconstruction regions, and overall enhanced detection efficiency of high-energy $\gamma$-rays. As a result, the statistics for high-energy bremsstrahlung $\gamma$-rays were significantly increased. Leveraging the unprecedented statistical precision and improved measurement accuracy of this experiment,  the HMT fraction in the nucleon momentum distribution of $^{124}$Sn nuclei has been determined quantitatively with improved precision \cite{Xu:2025mvv}. This result marks the first statistically high-precision extraction of the SRC fraction in low-energy HICs, achieved using bremsstrahlung $\gamma$-rays as a clean and sensitive probe.

In this paper, we present a comprehensive analysis of the SRC fraction extracted from the bremsstrahlung $\gamma$ spectrum in the 25 MeV/u \snsn~ experiment. Compared to the short article  reporting the primary physics outcome~\cite{Xu:2025mvv}, the present work emphasizes a systematic validation of the analysis and a transparent documentation of the methodologies employed. Specifically, we describe the detector setup, calibration procedures, and data analysis workflow in detail, and assess the robustness of the extracted SRC fraction through multiple independent and complementary approaches. On the theoretical side, a complete formulation of the IBUU-MDI transport model is presented together with systematic studies of parameter dependencies, demonstrating that the HMT fraction plays a dominant role in shaping the high-energy bremsstrahlung $\gamma$ spectrum. The SRC fraction is independently extracted using different coincidence selections, which yield consistent results within uncertainties and are incorporated into the systematic error budget. In addition, the bremsstrahlung $\gamma$ spectrum is reconstructed using an independent unfolding approach, enabling a direct comparison with theoretical predictions without detector-response folding. These combined analyses establish the robustness and reliability of the extracted HMT fraction and extend significantly beyond the scope of Ref.~\cite{Xu:2025mvv}.

This paper is organized as follows: 
    Section \ref{ExpSetup} outlines the experimental setup.
    Section \ref{ibuumodel} describes the IBUU-MDI transport model and the corresponding theoretical studies.
    Section \ref{hmtresults} focuses on data uncertainties and the extraction of the HMT fraction from the measured $\gamma$ spectra, including validation using different coincidence selections.
    Section \ref{RLmethods} presents an independent reconstruction of the bremsstrahlung $\gamma$ spectrum and its direct comparison with model predictions.
    Finally, Section \ref{conclusionoutlook} concludes with a summary and outlook.
The detailed descriptions of the trigger scheme, detector calibration, event reconstruction procedures, and background evaluation are provided in the Supplementary Material.

\section{Experimental Setup} \label{ExpSetup}

\begin{figure}[hptb]
    \centering
    \includegraphics[width=0.95\linewidth]{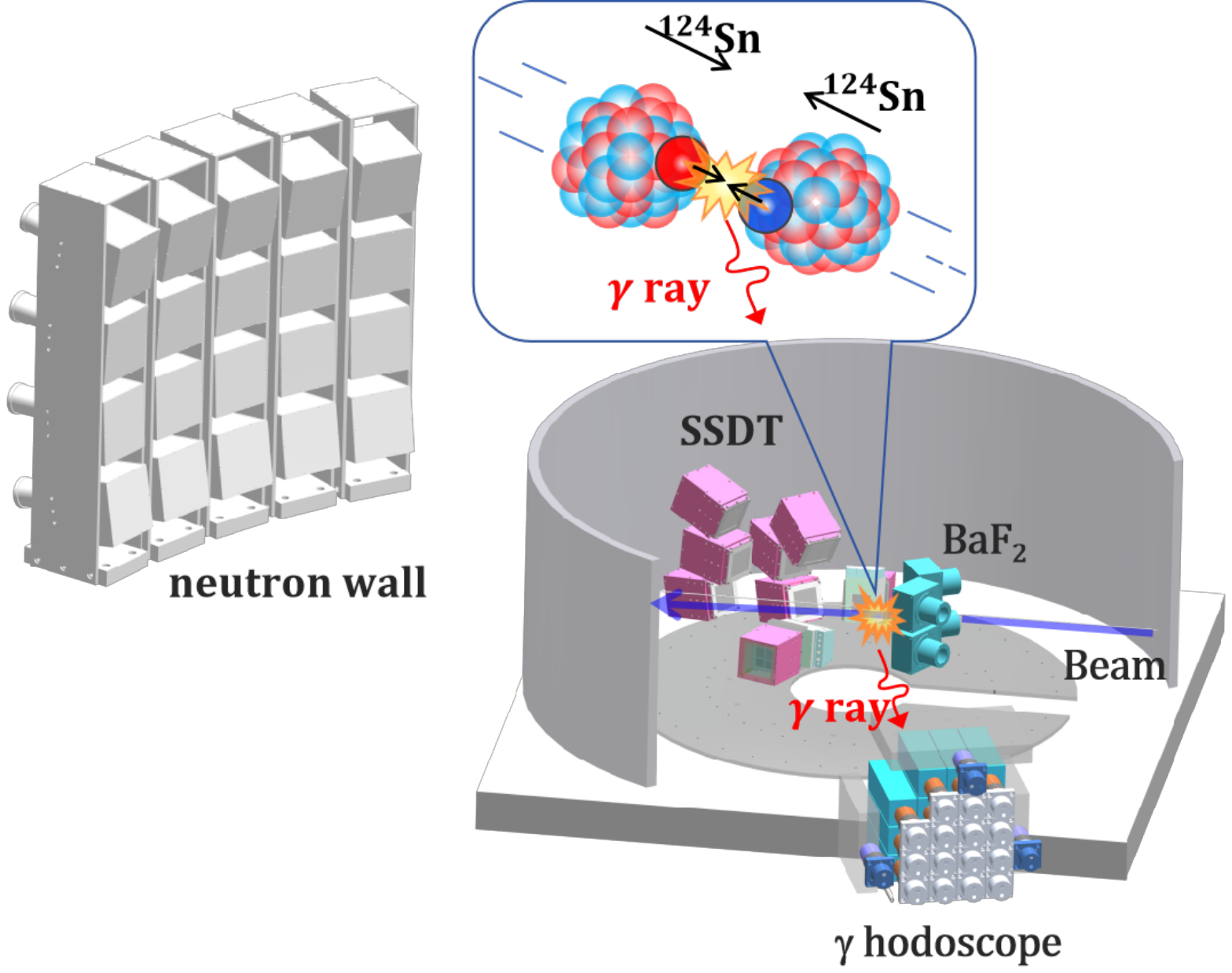}
    \caption{(Color Online) Experimental setup of CSHINE. The $\gamma$ hodoscope is located at $\theta_{\rm lab}=110^\circ$ to measure the bremsstrahlung $\gamma$-rays from the collisions of the $^{124}$Sn projectile on the $^{124}$Sn target.}
    \label{cshine}
\end{figure}

The experiment was performed at the Radioactive Ion Beam Line at Lanzhou (RIBLL-1).  The beam of \snnucl~ at 25 MeV/u was delivered by the Heavy Ion Research Facility at Lanzhou (HIRFL) and bombarded on a \snnucl~ target of 1 $\rm mg/cm^2$ installed in the chamber located at the final focal plane of RIBLL-1.  The products of the \snsn~ reactions were measured by the Compact Spectrometer for Heavy Ion Experiment (CSHINE), which is dedicated to the studies of heavy ion reactions at Fermi energies. The schematic view of the CSHINE setup in this experiment is presented in Fig. \ref{cshine}.  The light charged particles (LCPs) were detected by 8 silicon-strip detector telescopes (SSDTs), six of which provided good particle identification (PID). 
Each SSDT consists of a single-sided silicon strip detector (SSSD), a double-sided silicon strip detector (DSSD) and a $3\times3$ CsI(Tl) array. The overall energy resolution of each SSDT is better than 2\% \cite{Wang:2021jgu}, the pixel size of the SSDT is $4\times4$ mm. The charged isotopes for $Z \le 6$ can be clearly identified. The performance of the SSDTs had been demonstrated in previous experiments, one can refer to \cite{Guan:2021tbi,Wang:2021jgu,Guan:2021nfk} for details. 

The neutrons are measured by a plastic scintillator array consisting of $4\times 5$ units, covering the laboratory polar angle $17^\circ<\theta_{\rm lab}<53^\circ$ in partial azimuth. Each unit is a block of plastic scintillator with the size of $\rm 15\times 15\times 15 ~cm^3$ read out by a photomultiplier tube (PMT). The distance of the neutron array to the target is 200 cm. The neutron energies are measured by the time of flight (TOF) method. The start timing of TOF is provided by 4 $\rm BaF_2$ fast scintillators surrounding the target fired by the reaction $\gamma$-rays. For the performance of the neutron array, one can refer to \cite{Si:2024ujh}. Besides the neutron wall, a liquid scintillator neutron detector, with the sensitive volume of $\rm \phi 34.8 ~cm \times 102 ~cm$, was installed at 513 cm to the target at $\theta_{\rm lab}=60^\circ$. Two 8-inch PMTs are attached to the left and right ends of the sensitive volume.

The bremsstrahlung $\gamma$-rays of this analysis are measured by a CsI(Tl) hodoscope (CSHINE-Gamma), containing  15 units in a $4 \times 4$ configuration (with one corner being vacant). Each CsI(Tl) unit has the dimension of $\rm 7 \times 7 \times 25 ~cm^3$ and is coupled to a PMT of  Hamamatsu R2631 for signal readout. The radiation length of CsI(Tl) is $X_0 = 8.39 \rm ~g/cm^2$ and the Moliere radius is $r_{\rm M}=3.531 ~\rm cm$. The energy resolution of the units is about $3.6\%$ for 1 MeV $\gamma$-ray and about 2\% for $\gamma$-ray energy beyond 10 MeV. In order to  suppress the cosmic-ray muons, 3 thick plastic scintillators ($5 \times 30 \times 30 \rm ~cm^3$) were mounted surrounding the CsI(Tl) hodoscope on top, left and right sides. The linearity of the CsI(Tl) response to high-energy $\gamma$-rays was tested after the experiment using the quasi-monochromatic $\gamma$ beam at the Shanghai Laser Electron Gamma Source (SLEGS) \cite{XU2025170787}. The performance of the $\gamma$ hodoscope can be found in \cite{Qin:2022mzp}.

In order to determine the  centrality event-by-event, a small CsI(Tl) array was installed at the forward angle in the reaction chamber. The array is split into two modules and placed  on the left and right side of the beam, containing $4\times8$ units,  respectively. Each unit has the dimension of $1 \times 1 \times 5 \rm~ cm^3$ and is coupled to a PMT. This small CsI(Tl) array provides the multiplicity information of the charged particles (The $\gamma$-rays leave much lower energy deposits compared to the charged particles). This small CsI(Tl) array is not included in the trigger scheme. 

The trigger system of the \snsn~ experiment was designed to efficiently select events involving light charged particles, high-energy $\gamma$ rays, and neutrons. It was implemented using FPGA-based logic \cite{Guo:2022kwc} to combine timing signals from the SSDTs, the $\gamma$-ray hodoscope, the neutron array, and auxiliary scintillation detectors. Six trigger conditions were defined to cover different coincidence combinations, and a global trigger was issued when any condition was satisfied, ensuring both high efficiency and flexibility. For each event, the trigger type was encoded and stored, enabling event-by-event identification during offline analysis and forming the basis for subsequent coincidence and background studies. Detailed descriptions of the trigger logic and individual trigger conditions are provided in the Supplementary Material.

\section{IBUU-MDI Model Description}\label{ibuumodel}

\subsection{IBUU-MDI Model Overview}

The Isospin-dependent Boltzmann-Uehling-Uhlenbeck (IBUU) transport model \cite{Li:1996ix,Li:1997rc,Li:2003ts,Li:2018lpy,Bertsch1988rev} is employed in the theoretical calculations, which effectively describes the dynamics of nucleon-nucleon (NN) collisions, with its main equation given by
\begin{equation}
    \frac{\partial f}{\partial t}+\vec{v}\cdot \nabla_r f-\nabla_r U\cdot \nabla_p f=I_{\rm coll},
\end{equation}
where $f(\vec{r},\vec{p},t)$ is the probability of finding a particle at time $t$, with momentum $\vec{p}$ at position $\vec{r}$. $U$ represents the mean-field potential, and the evolution  of $f(\vec{r},\vec{p},t)$ in elastic and inelastic two-body collisions is governed by the collision term $I_{\rm{coll}}$,
\begin{equation}
\begin{aligned}
    I_{\rm{coll}} = & -\frac{1}{(2\pi)^3}\int \mathrm{d} ^{3}p_2 \mathrm{d} ^{3}p_{2^{'}} \mathrm{d}\Omega \frac{\mathrm{d}\sigma}{\mathrm{d}\Omega}v_{12} \\
& \times[ff_2(1-f_{1^{'}})(1-f_{2^{'}})-f_{1^{'}}f_{2^{'}}(1-f)(1-f_2)] \\
& \times \delta^3(\vec{p}_1+\vec{p}_2-\vec{p}_{1^{'}}-\vec{p}_{2^{'}}),
\end{aligned}
\end{equation}
where $\frac{\rm{d}\sigma}{\rm{d}\Omega}$ is the in-medium NN cross section and $v_{12}$ is the relative velocity of the two colliding nucleons. The coordinates $(\vec{r}_1,\vec{p}_{1})$ and $(\vec{r}_2,\vec{p}_{2})$ refer to the phase-space positions of nucleon 1 and nucleon 2 before collision, and change to $(\vec{r}_1,\vec{p}_{1^{'}})$ and $(\vec{r}_2,\vec{p}_{2^{'}})$ after collision. The Pauli blocking effect is also taken into account. The scattering is allowed only if the phase-spaces around $(\vec{r}_1,\vec{p}_{1^{'}})$ and $(\vec{r}_2,\vec{p}_{2^{'}})$ are unoccupied; otherwise, the scattering is suppressed \cite{Bertsch1988rev}. This effect is embodied in the last term of $I_{\rm{coll}}$. 

Moreover, one of the most important inputs in the IBUU model is the the mean-field potential. We adopt in this work the momentum-dependent interaction (MDI) potential derived from the Gogny effective interaction \cite{Das:2002fr},
\begin{equation}
\begin{aligned}
    U( \rho ,\delta ,\vec{p} ,\tau ) = & A_{u} (x) \frac{\rho _{\tau '}}{\rho _{0}} + A_{l} (x) \frac{\rho _{\tau }}{\rho _{0}} \\
  & +B( \frac{\rho}{\rho _{0}} )^{\sigma } (1-x\delta ^{2}) -8x\tau \frac{B}{\sigma +1} \frac{\rho ^{\sigma -1}}{\rho ^{\sigma }_{0}} \delta \rho _{\tau '}\\
  & +\frac{2C_{\tau ,\tau }}{\rho _{0}} \int \mathrm{d} ^{3} \vec{p'} \frac{f_{\tau }( \vec{r} ,\vec{p'} ) }{1+( \vec{p} -\vec{p'} ) ^{2} /\Lambda ^{2} } \\
  & +\frac{2C_{\tau ,\tau '}}{\rho _{0}} \int \mathrm{d} ^{3} \vec{p'} \frac{f_{\tau '}( \vec{r} ,\vec{p'} ) }{1+( \vec{p} -\vec{p'} ) ^{2} /\Lambda ^{2} },
\end{aligned}
\end{equation}
where $\tau=\pm1/2$ denotes the isospin ($+1/2$ for neutron and $-1/2$ for proton). $\rho_{n}$ and $\rho_{p}$ are the neutron and proton densities. $\rho=\rho_{n}+\rho_{p}$ is the total nucleon density. $\rho_0=0.16\,\rm{fm}^{-3}$ is the saturation density, and $\delta=(\rho_{n}-\rho_{p})/(\rho_{n}+\rho_{p})$ is the isospin asymmetry. The parameters $A_{u}(x)$, $A_{l}(x)$, $B$, $C_{\tau,\tau}$, $C_{\tau,\tau'}$, $\sigma$ and $\Lambda$ are taken from Ref.\cite{Das:2002fr}. Notably, the different choices of $x$ parameter correspond to different density dependence of the symmetry energy.

For comparison, a momentum-independent soft Bertsch-Kruse-Das Gupta (SBKD) potential is also used in IBUU simulations \cite{Bertsch1984sbkd},
\begin{equation}
    U(\rho)=A(\frac{\rho}{\rho_0})+B(\frac{\rho}{\rho_0})^{\sigma},
\end{equation}
where $A$, $B$ and $\sigma$ can be expressed in terms of the nuclear incompressibility coefficient in the equation of state \cite{Li1995prc2037}.

Another important input in the transport model is the NN elastic cross section. In IBUU, the isospin-dependent in-medium NN cross section using nucleon effective mass is given by \cite{Li2005sig}
\begin{equation}
\sigma^{\rm medium}_{\rm NN}=\sigma^{\rm free}_{\rm NN}(\frac{\mu_{\rm NN}^{\ast}}{\mu_{\rm NN}})^2,
\end{equation}
where $\mu_{\rm NN}$ and $\mu_{\rm NN}^{\ast}$ are the free-space reduced mass and in-medium reduced mass of the colliding nucleon pair, respectively. $\sigma^{\rm free}_{\rm NN}$ is the free-space NN cross section taken from experimental data \cite{Yong2011sig}.

The IBUU simulations utilize the initial single-nucleon momentum distribution $n(k)$ that incorporates the SRC and HMT effects \cite{Li:2018lpy,Xue:2016udl,Wang2017fnp}. For the symmetric nuclear matter (SNM), the HMT induced by SRC \cite{Hen:2014yfa} is given by
\begin{equation}
n_{\rm SNM}(k)=\left\{
\begin{array}{rcl}
A, & & {k\le k_{\rm F}}\\
C/k^4, & & {k_{\rm F}<k\le \lambda k_{\rm F}}\\
0, & & {k>\lambda k_{\rm F}},
\end{array} \right.
\end{equation}
where $k$ represents the single nucleon momentum. $k_{\rm F}$ is Fermi momentum and $\lambda$ is the high-momentum cutoff \cite{Xue:2016udl,Wang2017fnp}. The parameters $A$ and $C$ are determined by the normalization condition,
\begin{equation}
\begin{aligned}
\begin{cases}
4\pi \int\limits_{0}^{\infty}n_{\rm SNM}(k)k^{2}dk=1,\\
4\pi \int\limits_{k_{\rm F}}^{\infty}n_{\rm SNM}(k)k^{2}dk=R_{\rm HMT}.
\end{cases}
\end{aligned}
\end{equation}

For the asymmetric nuclear matter (ANM), the high-momentum component is linearly dependent on the isospin asymmetry \cite{PhysRevC.79.064308,PhysRevC.87.014314,PhysRevC.93.014619}. The $n(k)$ for ANM is parameterized as
\begin{equation}
n_{\rm ANM}(k)_p=\left\{
\begin{array}{rcl}
B, & & {k\le k^p_{\rm F}}\\
C(1+\delta)/k^4, & & {k^p_{\rm F}<k\le \lambda k^p_{\rm F}}\\
0, & & {k>\lambda k^p_{\rm F}},
\end{array} \right.
\end{equation}
\begin{equation}
n_{\rm ANM}(k){_n}=\left\{
\begin{array}{rcl}
B{'}, & & {k\le k^n_{\rm F}}\\
C(1-\delta)/k^4, & & {k^n_{\rm F}<k\le \lambda k^n_{\rm F}}\\
0, & & {k>\lambda k^n_{\rm F}},
\end{array} \right.
\end{equation}
where the parameters $B$ and $B{'}$ are determined by the normalization condition,
\begin{equation}
4\pi \int\limits_{0}^{\infty}n_{\rm ANM}(k)_{p(n)}k^{2}dk=1.
\end{equation}

The bremsstrahlung reaction channel is incorporated into the collision processes within the IBUU model, which employs the Bertsch criterion \cite{TMEP:2022xjg} and Pauli exclusion principle to determine whether a NN collision occurs. 
Bremsstrahlung photons are generated through two channels, the neutron-proton collisions ($np\rightarrow np\gamma$) and the proton-proton collisions ($pp \rightarrow pp\gamma$). However, the $np\gamma$ channel dominates as the lowest possible multipole radiation in the $pp$ collisions is quadrupole, whose intensity is much lower compared to the dipole radiation in $np$ collisions \cite{Cassing1990rev,Nifenecker1985mr}. Hence, we consider only the $np\gamma$ process in our calculations.

\begin{figure}[htb]
    \begin{minipage}{0.64\linewidth}
    \centerline{
	\includegraphics[width=\linewidth]{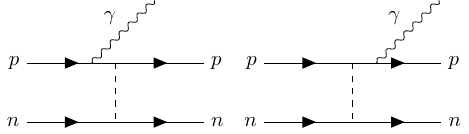}}
    \captionsetup{justification=centering}
    \caption*{(a)}
    \end{minipage}
    \hfill
    \begin{minipage}{0.34\linewidth}
    \centerline{
	\includegraphics[width=\linewidth]{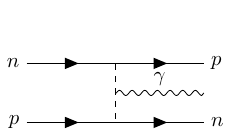}}
    \captionsetup{justification=centering}
    \caption*{(b)}
    \end{minipage}
    \caption{Feynman diagrams of bremsstrahlung photons in $np$ scattering from (a) external lines and (b) internal lines.}
    \label{FD}
\end{figure}

The bremsstrahlung photons emitted in the $np$ collisions are produced in the processes of deceleration and acceleration of the protons interacting with the neutrons. In the intermediate energy range, we adopt the neutral scalar $\sigma$ meson exchange model \cite{Cassing1990rev} in the present work, in which photons are emitted via the external processes, illustrated by the Feynman diagrams in Fig.\ref{FD} (a). At higher energies, one will have to take the charged meson exchange processes into consideration. The photons can also be emitted from the internal mesons, to which the electromagnetic field also couples \cite{Cassing1990rev,Schafer1991cme}, and the Feynman diagram is given by Fig.\ref{FD} (b).

Due to the low probability of producing bremsstrahlung photons in the reaction, the impact of bremsstrahlung on nucleon kinematics is also negligible, and hence the perturbation method is applied to calculate the probability of photon production in each $np$ collision. Based on the $\sigma$ meson exchange model mentioned above, a well fitted expression for the probability of elementary double differential photon production can be applied in IBUU simulations \cite{Gan1994298},
\begin{equation}
    \frac{{\rm d^2}P}{{\rm d}\Omega {\rm d}E_{\gamma}} = 1.671\times 10^{-7}\frac{[1-(E_{\gamma}/E_{\rm max})^2]^{\alpha}}{E_{\gamma}/E_{\rm max}},
\end{equation}
where $E_{\gamma}$ represents the energy of produced photons, and $E_{\rm max}$ represents the total available energy in the center-of-mass frame of the colliding neutron-proton pair. The coefficient $\alpha = 0.7319 - 0.5898\beta$, where $\beta$ represents the velocity of the nucleon. The total photon production probability per event is the sum of the probabilities of all $np$ collisions producing photons throughout the entire process.

\subsection{Sensitivity to Model Parameters}

In the IBUU simulations, one has to choose several important inputs which may influence the results of the final observables. Here we systematically check the sensitivity of bremsstrahlung photon production on the choices of parameters, namely the impact parameter, the nuclear mean-field potential, and the symmetry energy parameter. 

\begin{figure}[htbp]
	\centering
    \includegraphics[width=0.9\linewidth]{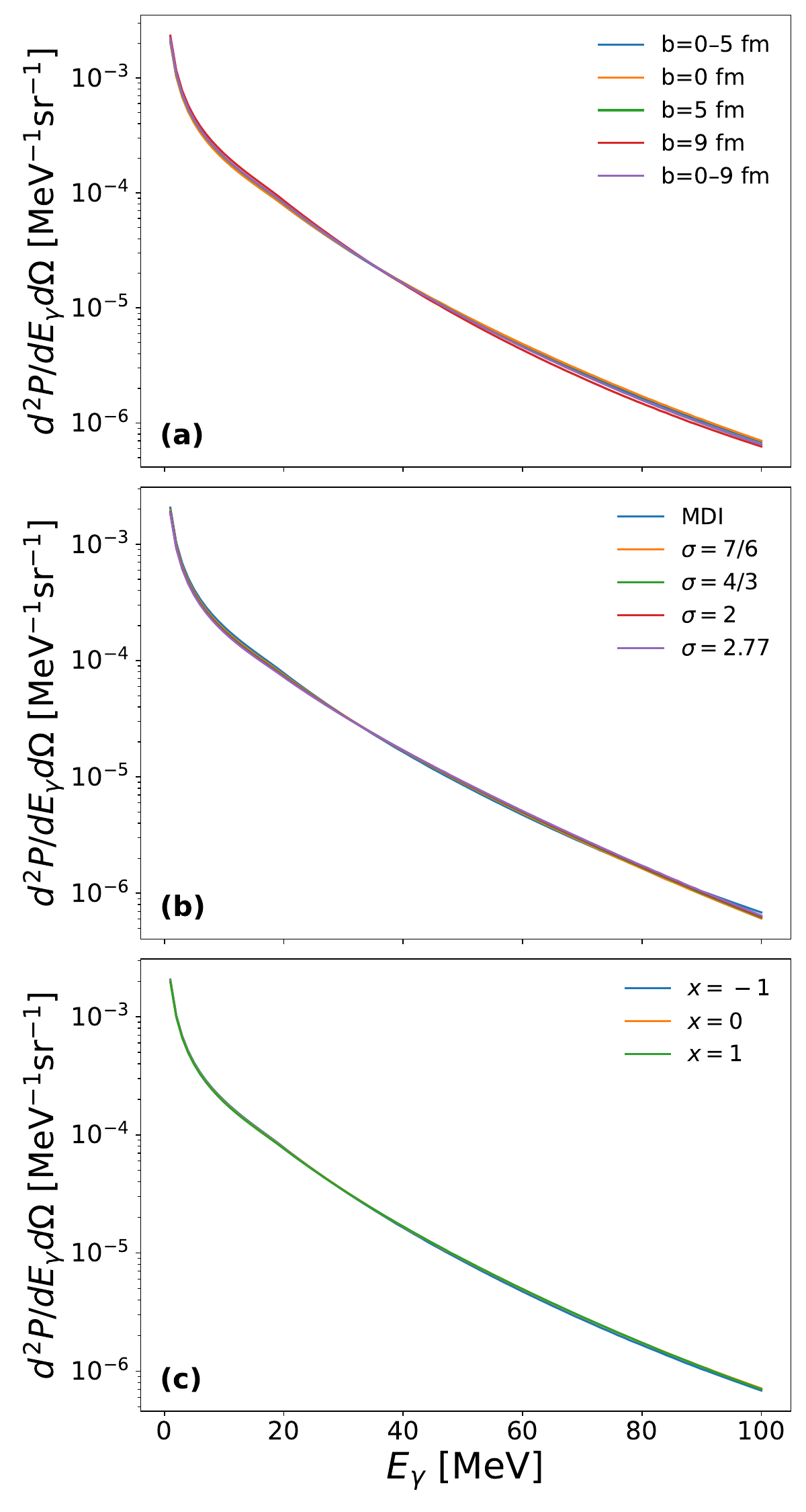}
    \caption{(a) The double differential photon production calculated by varying impact parameters. The impact parameters are taken to be $b=0\sim5\,\rm{fm}$, $b=0\sim9\,\rm{fm}$ and fixed values $b=0, 5, 9\,\rm{fm}$. The results using $b=0\sim9, 0, 5, 9\,\rm{fm}$ are normalized to the $b=0\sim5\,\rm{fm}$ case. (b) The double differential photon production calculated by varying the choice of nuclear mean-field potentials. We compare the results from the MDI potential  with those from the momentum-independent SBKD potential. The results from the SBKD potential are normalized to the MDI results. (c) The double differential photon production calculated by varying symmetry energy parameters. Three different $x$ parameters are employed in the simulations, namely $x=-1$, $x=0$ and $x=1$. The results using $x=0$ and $x=1$ are normalized to the $x=-1$ case.}
    \label{fig:combined_parameters}
\end{figure}

Firstly, we show in Fig.~\ref{fig:combined_parameters}(a) the variations of bremsstrahlung photon production by using different impact parameters. We adopt $b=0\sim5\,\rm{fm}$ in our IBUU simulations. To check the sensitivity of photon production on $b$ parameter, several other values are used, namely, $b=0\sim9\,\rm{fm}$ and $b=0, 5, 9\,\rm{fm}$. Their corresponding spectra are normalized to the $b=0\sim5\,\rm{fm}$ one. This is because we focus on the spectral shape similarities rather than absolute normalization. Although the absolute values of bremsstrahlung photon production spectra vary with different choices of impact parameters, but the spectral shape remains almost the same to the impact parameters, as seen in Fig.~\ref{fig:combined_parameters}(a). 

% \begin{figure}[htb]
% 	\centering
%     \includegraphics[width=0.9\linewidth]{./figures/b.pdf}
%     \caption{The double differential photon production calculated by varying impact parameters. The impact parameters are taken to be $b=0\sim5\,\rm{fm}$, $b=0\sim9\,\rm{fm}$ and fixed values $b=0, 5, 9\,\rm{fm}$. The results using $b=0\sim9, 0, 5, 9\,\rm{fm}$ are normalized to the $b=0\sim5\,\rm{fm}$ case.}
%     \label{b}
% \end{figure}

In Fig.~\ref{fig:combined_parameters}(b), the variations of bremsstrahlung photon production by using different nuclear mean-field potentials are shown. We adopt the MDI potential with the symmetry energy parameter $x=-1$ \cite{Qin:2023qcn}. The momentum-independent SBKD potential is used to check the sensitivity of mean-field potentials, and the parameters used in SBKD potential are listed in Table\ref{Table1}. The $\sigma$ parameter represents the softness or stiffness of the potential with the variation of the nuclear density \cite{Bertsch1984sbkd}. For the possible influence of symmetry energy parameters in MDI, we adopt three different parameters $x=-1$, $x=0$ and $x=1$ in simulations (see Fig.~\ref{fig:combined_parameters}(c)). The tests have shown that within the energy region investigated in this work, our conclusions are not affected by the choice of nuclear potentials or the symmetry energy parameters. This is probably because the bremsstrahlung photons interact with nucleons only electromagnetically, and are not sensitive to the isoscalar and isovector parts of nuclear interactions.

% \begin{figure}[htb]
% 	\centering
% \includegraphics[width=0.9\linewidth]{./figures/potential.pdf}
%     \caption{The double differential photon production calculated by varying the choice of nuclear mean-field potentials. We compare the results from the MDI potential  with those from the momentum-independent SBKD potential. The results from the SBKD potential are normalized to the MDI results.}
%     \label{pot}
% \end{figure}

\begin{table}[hbtp]
    \centering
    \caption{The parameters $\sigma$, $A$ (MeV) and $B$ (MeV) used in the momentum-independent SBKD potential and their corresponding incompressibility coefficients $K$ (MeV) in the equation of state of nuclear matter.}
    \label{Table1}
    \begin{tabular}{ccccc}% 
    \toprule 
    $\sigma$ & $A$ (MeV)   & $B$ (MeV) & $K$ (MeV) \\ 
    \midrule %[2pt]  
    7/6 & -356 & 303 & 201  \\
    4/3 & -218 & 164 & 237  \\
    2   & -124 & 70.5 & 377\\ 
    2.77   & -103.22 & 49.56 &540 \\ 
    \bottomrule %[2pt]     
    \end{tabular}
\end{table}

% \begin{figure}[htb]
% 	\centering
% \includegraphics[width=0.9\linewidth]{./figures/x.pdf}
%     \caption{The double differential photon production calculated by varying symmetry energy parameters. Three different $x$ parameters are employed in the simulations, namely $x=-1$, $x=0$ and $x=1$. The results using $x=0$ and $x=1$ are normalized to the $x=-1$ case.}
%     \label{x}
% \end{figure}

\section{Experimental $\gamma$-Ray Spectra}\label{hmtresults}

\subsection{Analysis of Systematic Uncertainty}

The systematic uncertainties primarily stem from three sources: (1) variations in the spectra obtained from radioactive source calibrations on different dates during the experiment, (2) the choice of normalization energy range for high-energy cosmic rays during background subtraction, and (3) the linear assumption of the detector response at higher $\gamma$ energy levels. Analysis shows that the choice of normalization energy ranges exhibits no significant variations, indicating that the effect of cosmic ray normalization range on the $\gamma$-ray spectrum is negligible.

The experimental data included three sets of linear calibrations obtained from radioactive sources on different dates. Among these, the $\gamma$ energy spectrum calibrated using the dataset with the highest statistics was designated as the central spectrum. In addition to the linear calibration, we introduced two alternative detector response correction functions (DRCFs) characterizing the differences between the linear calibration and the actual quadratic calibration obtained on SLEGS test experiment \cite{XU2025170787}. Then a total of nine distinct $\gamma$ energy spectra can be derived. Excluding the central spectrum, we calculated the standard deviation of the remaining eight spectra for each bin as the systematic uncertainty. During the evaluation of the likelihood functions and $\chi^2$ (see next), the values derived from the central spectrum were used as the central points. The likelihood function and $\chi^2$ values computed from the remaining spectra were utilized to determine the associated errors (standard deviations). This methodology enabled us to obtain likelihood and $\chi^2$ functions with systematic uncertainties at different $R_{\text{HMT}}$ values.

\subsection{Experimental $\gamma$ spectrum in slow coincidence}

\begin{figure}[hptb]
     \centering
    \includegraphics[width=0.9\linewidth]{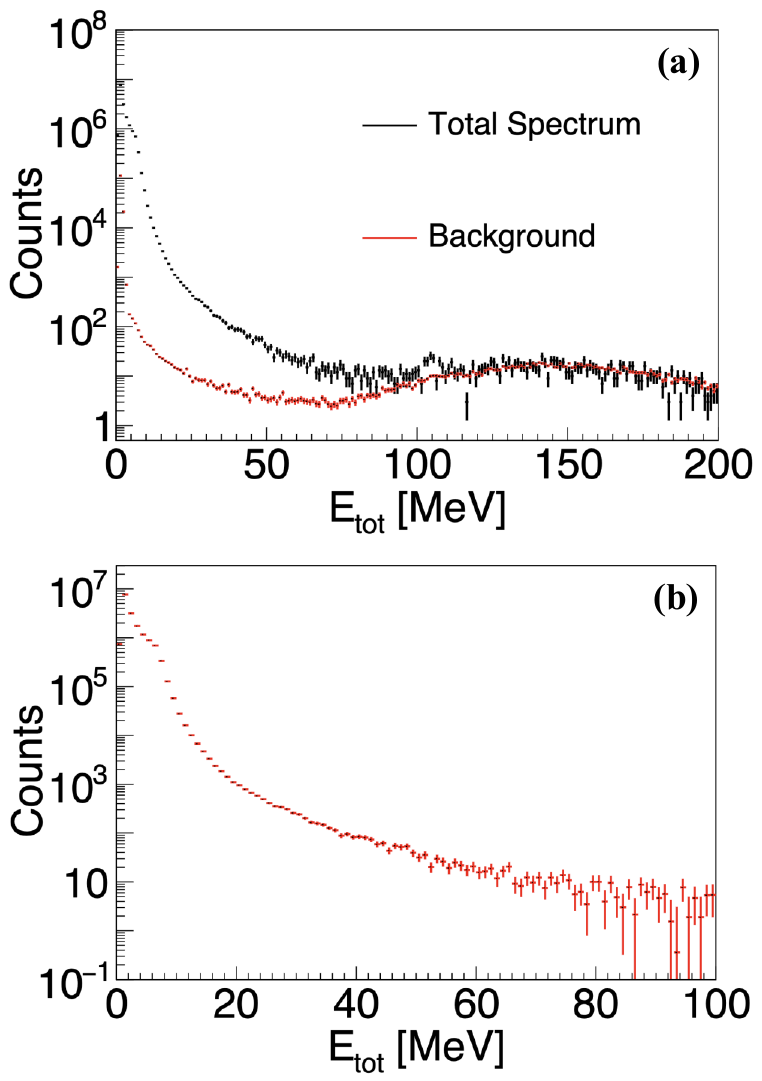}
    \caption{(Color Online) (a) Total energy Spectrum of beam-on (black) in all trigger conditions and beam-off measurement (red).  (b) The $\gamma$ energy spectrum after subtracting the background.}
    \label{EnergySpec_slow}
  \end{figure}

As shown in the primary analysis, the method introduced in the previous experiment \cite{Qin:2023qcn} is applied. All $\gamma$ events in the slow coincidence window are counted. Namely a valid timing of the reconstructed $\gamma$ is found within the TDC range. The background is obtained by the beam-off measurement. The experimental data calibration and background subtraction were performed using the background data and radioactive source calibration data on March 8, 2024. Calibration files were generated based on these data and subsequently applied to all experimental and background data for calibration and reconstruction. 
This process yielded the total beam-on spectrum and the beam-of background spectrum, shown as the black and red lines in Fig.~\ref{EnergySpec_slow}(a), respectively. 
The red background spectrum was scaled by a normalization factor $r_{\rm n}$. Specifically, the total counts above $110 ~\rm MeV$ in both the total spectrum ($Y_{\rm on}$) and the background spectrum ($Y_{\rm off}$) were calculated, and the yield ratio is defined as $r_{\rm n}=\frac{Y_{\rm on}(E_{1})}{Y_{\rm off}(E_{1})}$, where the subscript `on' and `off' represent the beam-on and beam-off spectrum, respectively. $E_1=110$ MeV is the low-energy border taken to do the normalization, and the upper border is set to 200 MeV. Then the ratio $r_{\rm n}$ is applied to the entire background spectrum, resulting in the red histogram in Fig.~\ref{EnergySpec_slow}(a).  By subtracting the scaled background spectrum from the total energy spectrum, the background-subtracted  $\gamma$ spectrum was obtained, as shown in Fig.~\ref{EnergySpec_slow}(b), where we take $0-100$ MeV as the energy range of the total $\gamma$-rays.

\begin{figure}[hptb]
    \centering
    \includegraphics[width=0.95\linewidth]{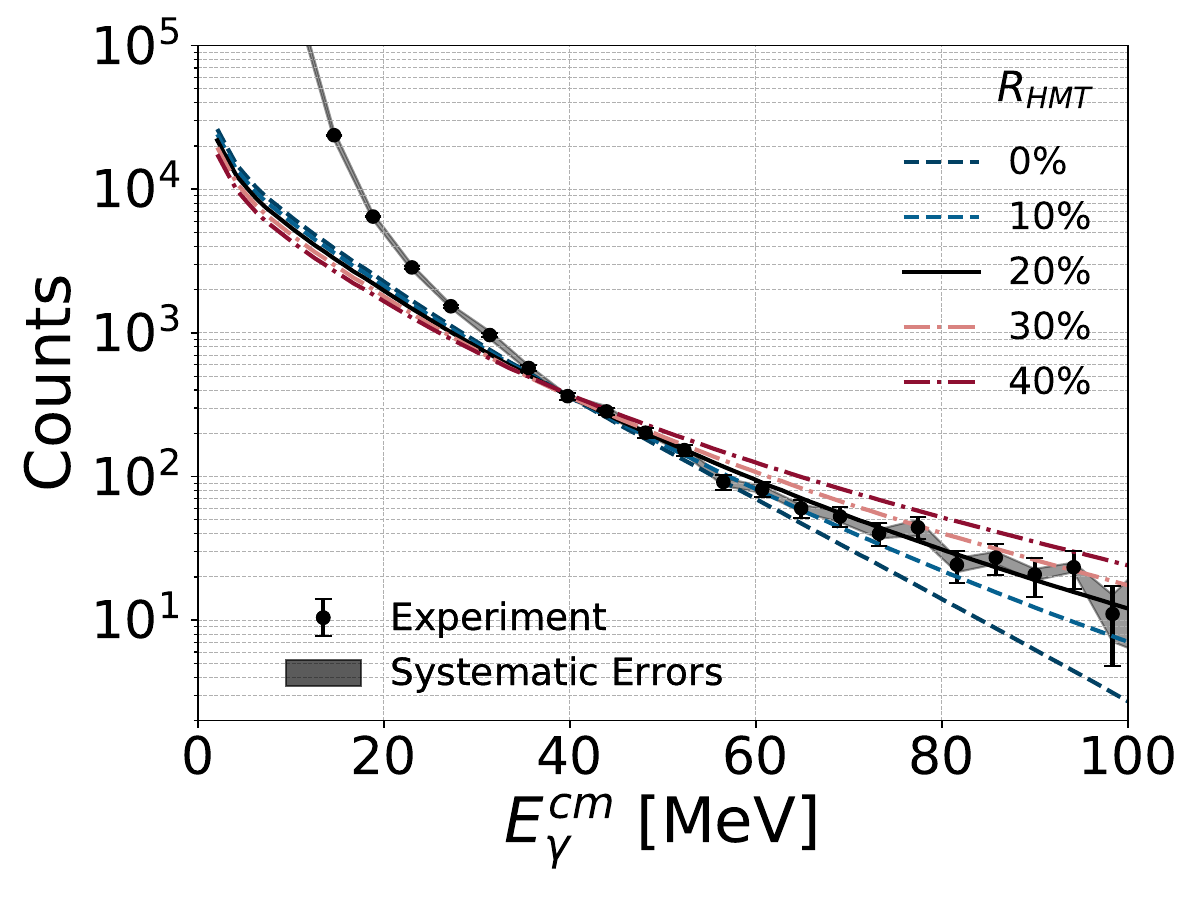}
    \caption{(Color Online) Comparison of the rebinned experimental $\gamma$ spectrum (black dots) with statistical uncertainties (error bars) and systematic uncertainties (gray shaded areas) in the c.m.  frame. Several key theoretical curves with different $R_{\rm HMT}$, processed with the detector filter, are overlaid for comparison.  }
    \label{BUUandexppy}
\end{figure}

Fig.~\ref{BUUandexppy} illustrates the rebinned experimental  $\gamma$  spectrum in the c.m. frame. The black dots represent the central spectrum, with statistical uncertainties indicated by error bars, and the gray shaded regions denote the systematic uncertainties at each energy point. Several key theoretical curves, processed with the detector filter, are presented for comparison. Unlike in our previous measurement \cite{Qin:2023qcn, Xu:2024oct}, in this experiment, the  $\gamma$  hodoscope was operated in active triggering mode, preventing the determination of the number of $np$ collisions in heavy-ion reactions as a normalization factor for the  $\gamma$  spectrum. Consequently, the comparison between the experimental data and theoretical curves predicted by the IBUU model focused on spectral shape similarities rather than absolute normalization. To align the range of theoretical curves with the experimental spectrum, the theoretical values were uniformly scaled by a factor of $2.5 \times 10^9$, which had no impact on determining the $R_{\rm HMT}$, as it did not affect the comparison of the shape differences between the experimental and model-predicted spectra.

Energies below $30~\rm MeV$  are primarily influenced by collective resonance and statistical emissions, as well as the random coincidence. Therefore, we selected an energy range of $35 < E_{\gamma}^{\rm cm} < 100~\rm MeV$ as the central analysis interval and calculated the likelihood function between the experimental $\gamma$ spectrum at each point within this range and various IBUU theoretical curves.

The comparison between the IBUU model and the experimental data was performed using the maximum likelihood analysis method, which assumes that the total counts are fixed and that the experimental spectrum can be treated as a histogram, with each bin following a multinomial distribution. 
The likelihood function can be defined as follows, primarily focusing on the similarity in the shape of the spectra,
\begin{equation}
    \begin{aligned}
        L(R_{\text{HMT}}) &= n!\prod_{i}^{\text{range}}\frac{1}{n_i!}p_i^{n_i}(R_{\text{HMT}}),\\
        \ln L(R_{\text{HMT}}) &= \sum_{i}^{\text{range}}n_i\ln p_i(R_{\text{HMT}})-\sum_{i}^{\text{range}}\ln n_i!+\ln n!,\\
    \end{aligned}
\end{equation}
where $i$ represents the sum of experimental points within a certain statistical analysis interval and $n_i$ is the count of $i^{\rm th}$ experimental data point.
$p_i$ represents  the probability that the theoretical model predictions falling within the corresponding histogram bin of the experimental data under the specified statistical analysis interval and for a given $R_{\text{HMT}}$.
Additionally, in the given statistical interval, all the $p_i$ should be normalized.
The second and third terms are model independent, which can be neglected to simplify the calculation.  Thus, a simplified logarithmic likelihood function can be defined as
\begin{equation}
    \ln L'(R_{\text{HMT}}) = \sum_{i}^{\text{range}}n_i\ln p_i(R_{\text{HMT}}).
\end{equation}

\begin{figure}[hptb]
    \centering
    \includegraphics[width=0.95\linewidth]{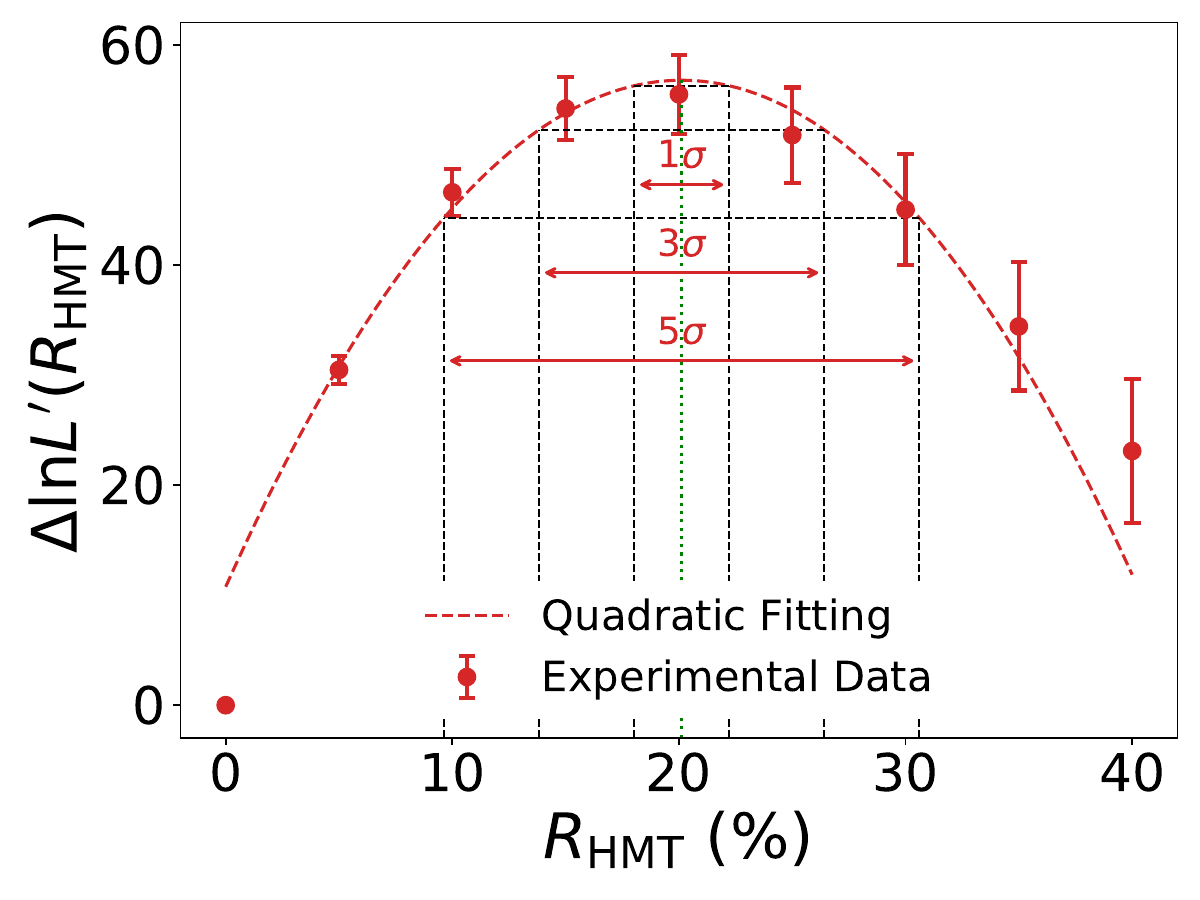}
    \caption{The likelyhood function values of different $R_{\text{HMT}}$ and the corresponding quadratic fitting in the energy range of $35~\rm MeV$ to $100~\rm MeV$.}
    \label{HMTComparisons}
\end{figure}

The results are presented as red dots with error bars in Fig.~\ref{HMTComparisons}, where the central values are derived from the central $\gamma$ spectrum. To standardize the likelihood function values obtained from different $\gamma$ spectra, the likelihood function value corresponding to the theoretical curve derived from $R_{\text{HMT}} = 0\%$  was selected as the reference point ($\ln L(R_{\text{HMT}} = 0\%) = 0$). The relative likelihood function values, $\Delta\ln L'(R_{\text{HMT}})$, for all other points were calculated relative to this baseline in each spectrum. The error bars at each $R_{\text{HMT}}$ represent the standard deviations of the relative likelihood values obtained from various $\gamma$ spectra generated using different calibration parameters in comparison to the model predictions with different $R_{\rm HMT}$.

The likelihood function reaches its maximum near $R_{\text{HMT}} = 20\%$ and follows an approximately quadratic distribution. The trend was fitted with a quadratic function within the range of $5\% < R_{\text{HMT}} < 35\%$, depicted as the dashed curve in the figure, which also yielded a maximum value at $R_{\text{HMT}} = (20.1\pm 0.4)\%$, where the uncertainty of 0.4\% reflects the fitting error from the quadratic model.

To determine the confidence intervals for the HMT ratio based on the likelihood function distribution, $R_{\text{HMT}}$  values were identified that correspond to relative likelihood values smaller than the maximum by $\frac{m^2}{2}$, which represent confidence levels of  $\pm m\sigma$. The  $R_{\text{HMT}}$ ranges for likelihood values reduced by $0.5$, $4.5$ and $12.5$ from the quadratic maximum correspond to  $\pm 1\sigma$, $\pm 3 \sigma$ and $\pm 5 \sigma$  confidence intervals, respectively, as indicated by the black dashed lines in the figure. Based on the experimentally measured $\gamma$ spectrum, the value of $R_{\text{HMT}}$ is determined at the $1\sigma$ confidence level as
\begin{equation}
    R_{\text{HMT}} = (20.1\pm 2.1)\%.
\end{equation}

Worth mentioning, the spectrum obtained here does not show a broad hump-like structure near $E_{\gamma}\approx 60$ MeV, at variance with the $\gamma$ spectrum obtained in the previous Kr+Sn experiment at the same beam energy \cite{Qin:2023qcn}, where the statistics was much lower. According to the new result, the presence of $np \rightarrow d\gamma$ channel in the collision is not apparently supported.

\subsection{Experimental $\gamma$ spectrum in fast coincidence}

 Alternatively, one can analyze the $\gamma$ spectrum in a different approach as a cross check, where both the $\gamma$ spectrum and the background are extracted in beam-on experiment. Specifically, the $\gamma$ events are accumulated in the fast coincidence gate defined by the trigger, while the background is obtained in a time window with the same width but far away from the main trigger time window. By this approach, one ensures that only random coincidence is contained in the time window far away from the main trigger gate.
 
 Noticeably, however, the events triggered  by  the condition requiring two or more LCPs detected in the SSDTs (\ssdtwo) shall be excluded now, because the $\gamma$ time are spreading across the entire TDC range due to serious timing jitter of the trigger signal (see Supplementary Materials). Otherwise, the truly coincident $\gamma$ events are recorded in the background window and the subtraction leads to a wrong energy spectrum.

To isolate true $\gamma$ events, we exclude all events of the \ssdtwo~ trigger within the main peak region and define two time windows: a main window and a window well separated from the former, as specified below,

\begin{itemize}
    \item The events with $T_{\rm core}$ in the range $-350 ~\text{ns} < T_{ \rm core} < -50 ~\text{ns}$ are considered as true $\gamma$ events.
    \item The events in the range $50 ~\text{ns} < T_{ \rm core} < 350 ~\text{ns}$ are taken as background.
\end{itemize}

\begin{figure}[hptb]
     \centering
    \includegraphics[width=0.9\linewidth]{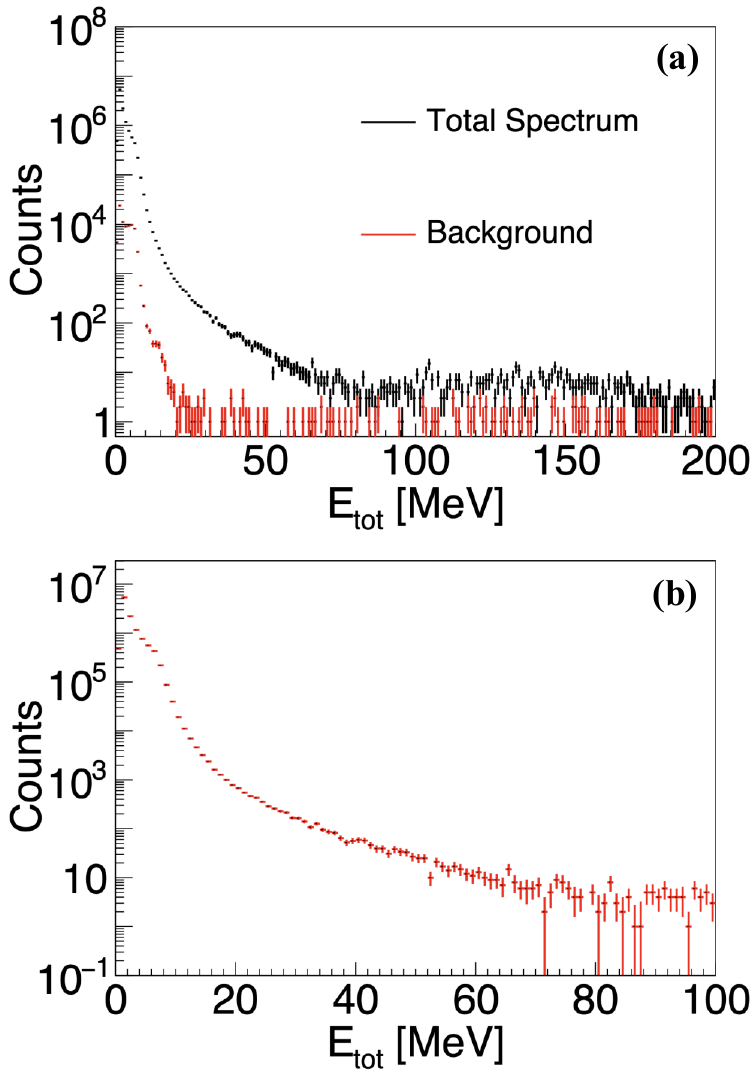}
    \caption{(Color Online) (a) Total energy Spectrum in the fast coincidence window  excluding \ssdtwo~ trigger condition (black) and in the accidental coincidence window (red).  (b) The $\gamma$ energy spectrum after subtracting the background.}
    \label{EnergySpec_fast}
  \end{figure}

The energy spectra constructed from these two time windows with equal width are shown in Fig.~\ref{EnergySpec_fast}(a). The black curve represents the $\gamma$-ray energy spectrum within the fast coincidence window, while the red curve corresponds to the background window. Panel (b) shows the resulting $\gamma$ energy spectrum after subtracting the background directly. Since the time width is the same, no more normalization is needed.

\begin{figure}[hptb]
     \centering
    \includegraphics[width=0.95\linewidth]{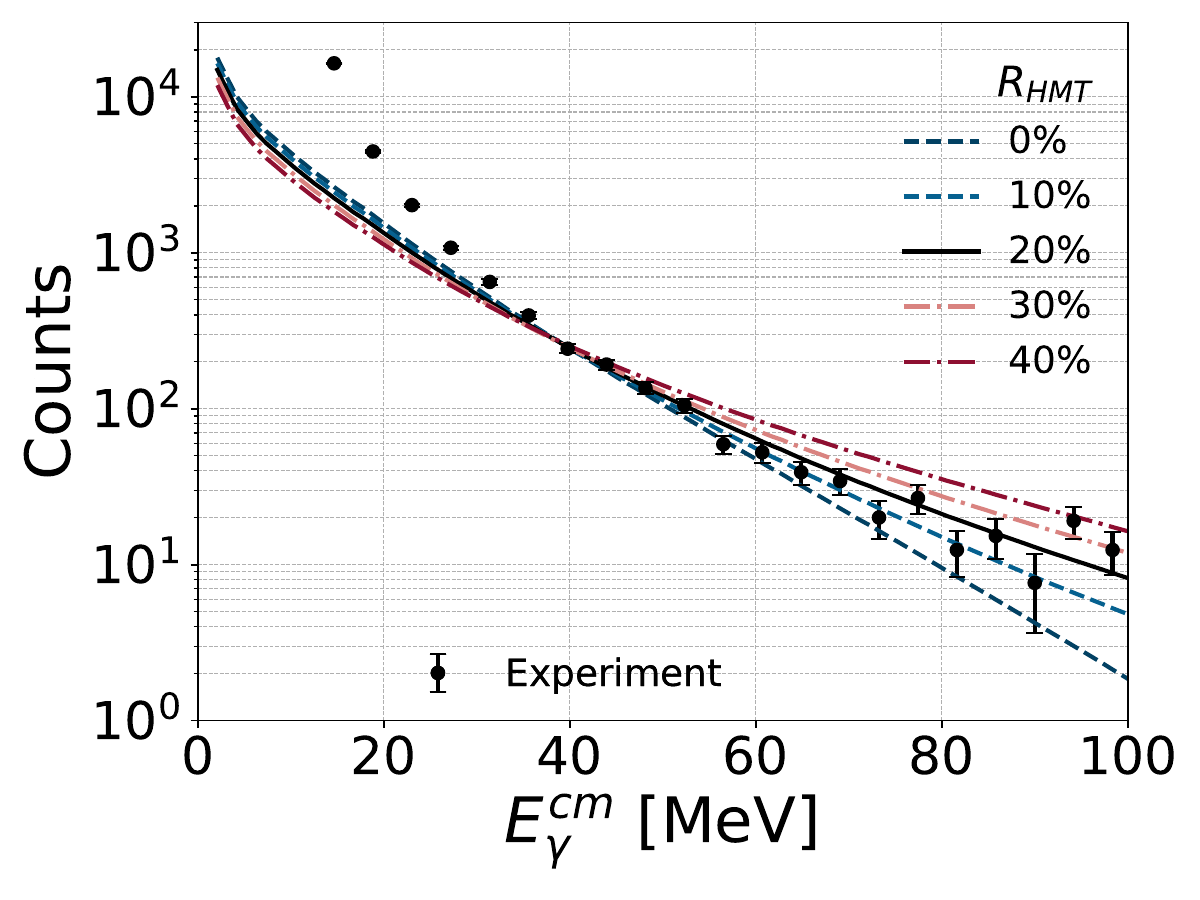}
    \caption{(Color Online) Comparison of the rebinned experimental $\gamma$ spectrum  excluding \ssdtwo~ trigger condition   with   key theoretical curves.}
    \label{compare_fast}
  \end{figure}

The same comparison between the background-subtracted $\gamma$ spectrum (excluding the \ssdtwo~ trigger) and theoretical model predictions, scaled by a factor of $1.7\times 10^9$, is shown in Fig.~\ref{compare_fast}. 
Using a likelihood-based method, we compare the experimental spectrum with IBUU model predictions and extract an optimal $R_{\text{HMT}}=(18.0\pm2.8)\%$. This value is consistent, within uncertainties, with the result obtained via the beam-off background subtraction method, indicating that the extracted SRC fraction is robust and independent of the specific background subtraction method.

Due to the reduced statistics of photons, particularly above 60 MeV, even slight fluctuations in the high-energy region may influence the final result. Thus, the 2\% deviation from the previous analysis is taken as a systematic uncertainty. Combined, the total uncertainty is $\sqrt{2.1^2+2^2+0.4^2}\approx3$. As the final result, the fraction of HMT is written as

\begin{equation}
    R_{\text{HMT}} = (20\pm 3)\%.
\label{rhmt_fnl}
\end{equation}

\section{Reconstruction of the Original $\gamma$ Spectrum}\label{RLmethods}

Finally, in order to offer a $\gamma$ spectrum that can be compared with other theoretic calculations without knowing the response matrix, we also tried to reconstruct the original $\gamma$ spectrum from the measured one by solving an inverse problem. In addition, the results of $R_{\rm HMT}$ can also be checked one more time from the original spectrum. For this purpose, the Richardson Lucy (RL) algorithm, which has been widely applied in nuclear physics \cite{Vargas:2013kga,Danielewicz:2021vqq,Nzabahimana:2022ndq,Nzabahimana:2023tab,Mamba:2024pch,Xu:2024dnd,Nzabahimana:2025qdj}, is adopted to solve the inverse problem \cite{Xu:2024oct}.

\subsection{Unfolding Method and Detector Response}

In optical deblurring problems, the true property $\mu$ of a photon is measured as $\nu$, with their distributions $ \mathcal{F}(\mu) $ and $f(\nu)$ related by:

\begin{equation}
f(\nu) = \int d\mu P(\nu|\mu) \mathcal{F}(\mu),
\end{equation}
where $ P(\nu|\mu) $ represents the conditional probability of measuring a photon with the real property $ \mu $ as $ \nu $.

Analogously, in HIC experiments, the true photon energy $\mu$ is detected as $\nu$ due to the detector filter. This allows the RL algorithm to be applied for reconstructing the original energy spectrum. In discretized form, the relationship between the distribution function of the real energy spectrum $\mathcal{E}(\mu)$ and the measured energy spectrum $e(\nu)$ can be related via the detector response matrix $D_{ij}$:
\begin{equation}
e_i = \sum_j D_{ij} \mathcal{E}_j,
\end{equation}
where $D_{ij}$ is obtained from Geant4 simulations with the given experimental setup and data analysis scheme. It quantifies the conditional probability that a photon with real energy $\mu$ will be detected as $\nu$.

The RL algorithm updates the estimated spectrum iteratively:
\begin{equation}
\mathcal{E}_i^{(r+1)} = A_i^{(r)} \cdot \mathcal{E}_i^{(r)},
\end{equation}
with the amplification factor $ A_i^{(r)} $
\begin{equation}
A_i^{(r)} = \sum_j \frac{e_j}{e_j^{(r)}} T_{ji},
\end{equation}
where the predicted measurement at the $r$-th iteration $ e^{(r)} $ is:
\begin{equation}
e_j^{(r)} = \sum_i D_{ji} \mathcal{E}_i^{(r)}.
\end{equation}
The normalized transformation matrix $T_{ji}$ is defined as
\begin{equation}
T_{ji} = \frac{W_j D_{ji}}{\sum_{j’} W_{j’} D_{j’i}},
\end{equation}
where $W_j = \sqrt{e_j}$ reflects Poisson statistics of the measured counts.

To assess statistical uncertainties, we performed multinomial sampling of the measured spectrum. Multiple pseudo-spectra are generated while preserving the total event count, and each is independently processed through the RL algorithm. The resulting ensemble of reconstructed spectra allows us to compute the standard deviation for each energy bin, which serves as our uncertainty estimate. In the RL reconstructed $\gamma$-ray spectrum, the central values are obtained by the measured spectrum, while the error bars reflect the propagated statistical uncertainties derived from the multinomial sampling. This method robustly accounts for statistical fluctuations in the measurement and their impact on the spectral reconstruction.

\subsection{Original $\gamma$-Ray Spectrum}

The initial values for all iterative solutions of different spectra are set to the theoretical curve with $R_{\text{HMT}} = 0\%$. 
This choice eliminates any prior bias toward a nonzero HMT ratio in the initial condition. 
By starting with $R_{\text{HMT}} = 0\%$, the iterative process avoids the risk of a pre-imposed HMT contribution influencing the reconstructed spectrum. 
Furthermore, this initialization improves both the efficiency and accuracy of the iterations, ensuring that the final solution is derived purely from the input data without relying on an arbitrary starting condition.

The number of iterations is a critical factor in ensuring the convergence of the RL algorithm while maintaining computational efficiency. To determine the optimal stopping point, we monitor the standard deviation $\delta_{\text{exp}}$ between the experimental spectrum and the predicted spectrum (calculated using the detector response matrix) after each iteration. The quantity $\delta_{\text{exp}}$ is defined as:

\begin{equation}
\delta_{\text{exp}} = \sum_i \left( \frac{e_i^P - e_i}{\sqrt{e_i}} \right)^2 / N_p,
\end{equation}
where $e_i$ represents the input measured value for the $i^{\rm th}$ energy bin, $e_i^P$ is the predicted value from the current iteration, and $N_p$ is the total number of bins. Here, $N_p = 100$, corresponding to the energy range of $1~\text{MeV}$ to $100~\text{MeV}$ with a step size of $1~\text{MeV}$.

The convergence condition is established by monitoring the relative change in $\delta_{\text{exp}}$ between consecutive iterations. The iteration is terminated when the rate of change of $\delta_{\text{exp}}$ becomes less than $1 \times 10^{-2}$, as expressed by:

\begin{equation}
\left| \frac{\delta_{\text{exp}}^{(r)} - \delta_{\text{exp}}^{(r-1)}}{\delta_{\text{exp}}^{(r-1)}} \right| < 1 \times 10^{-2}.
\end{equation}

This threshold is carefully chosen to strike a balance: avoiding premature termination caused by insufficient iterations and mitigating error amplification that may arise from excessive iterations.

\begin{figure}[hptb]
    \centering
    \includegraphics[width=0.9\linewidth]{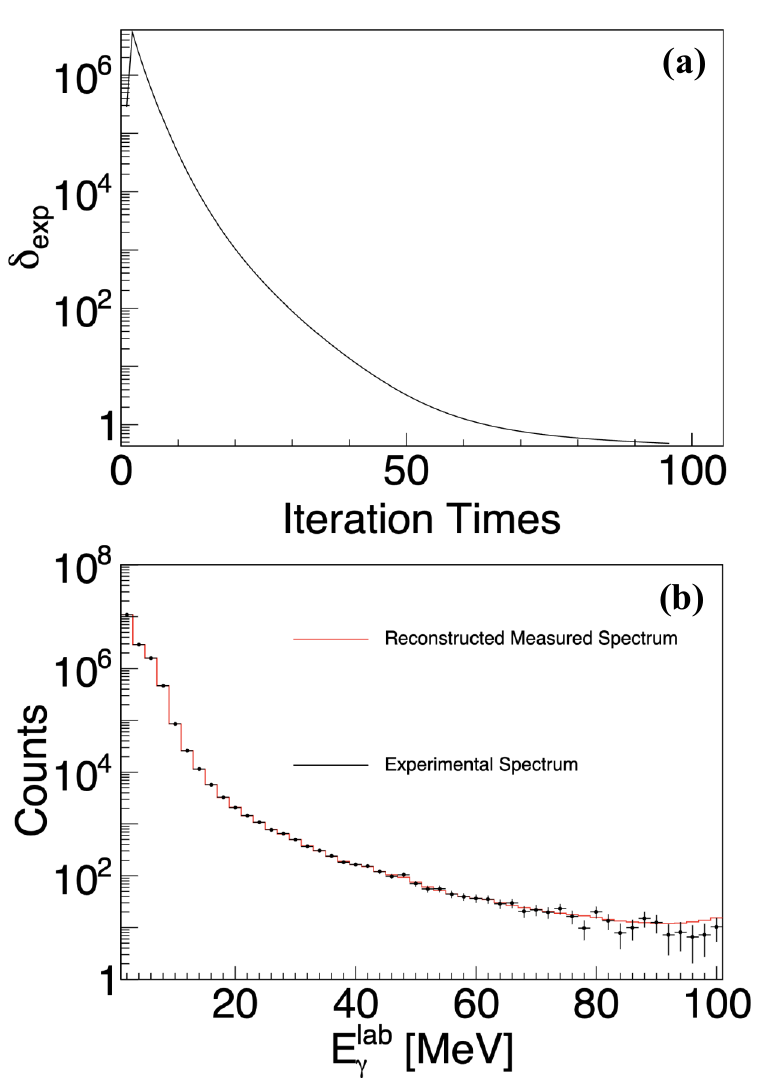}
    \caption{(a) The evolution of $\delta_{\text{exp}}$ as a function of iteration times in solving central spectrum.
    (b) The comparison between the reconstructed measured spectrum and the original measured spectrum in the experiment.}
    \label{IterationDisplay}
\end{figure}

The evolution of $\delta_{\text{exp}}$ as a function of iteration number in solving the central spectrum is shown in Fig.~\ref{IterationDisplay}(a). It can be observed that $\delta_{\text{exp}}$ gradually decreases with increasing iteration times, indicating a systematic improvement in the agreement between the experimental and predicted spectra. The end of the curve is at the convergence point, signaling the optimal termination of the iterative process.

To verify the reliability of the reconstruction, the reconstructed spectrum in the lab frame is folded back using the detector filter matrix to obtain the corresponding predicted measured spectrum. 
This reconstructed measured spectrum is compared with the original central measured spectrum, as shown in Fig.~\ref{IterationDisplay}(b). 
The close agreement between the two spectra demonstrates that the reconstructed real $\gamma$ spectrum is both accurate and efficient, validating the performance of the iterative algorithm.

\begin{figure}[hptb]
    \centering
    \includegraphics[width=0.95\linewidth]{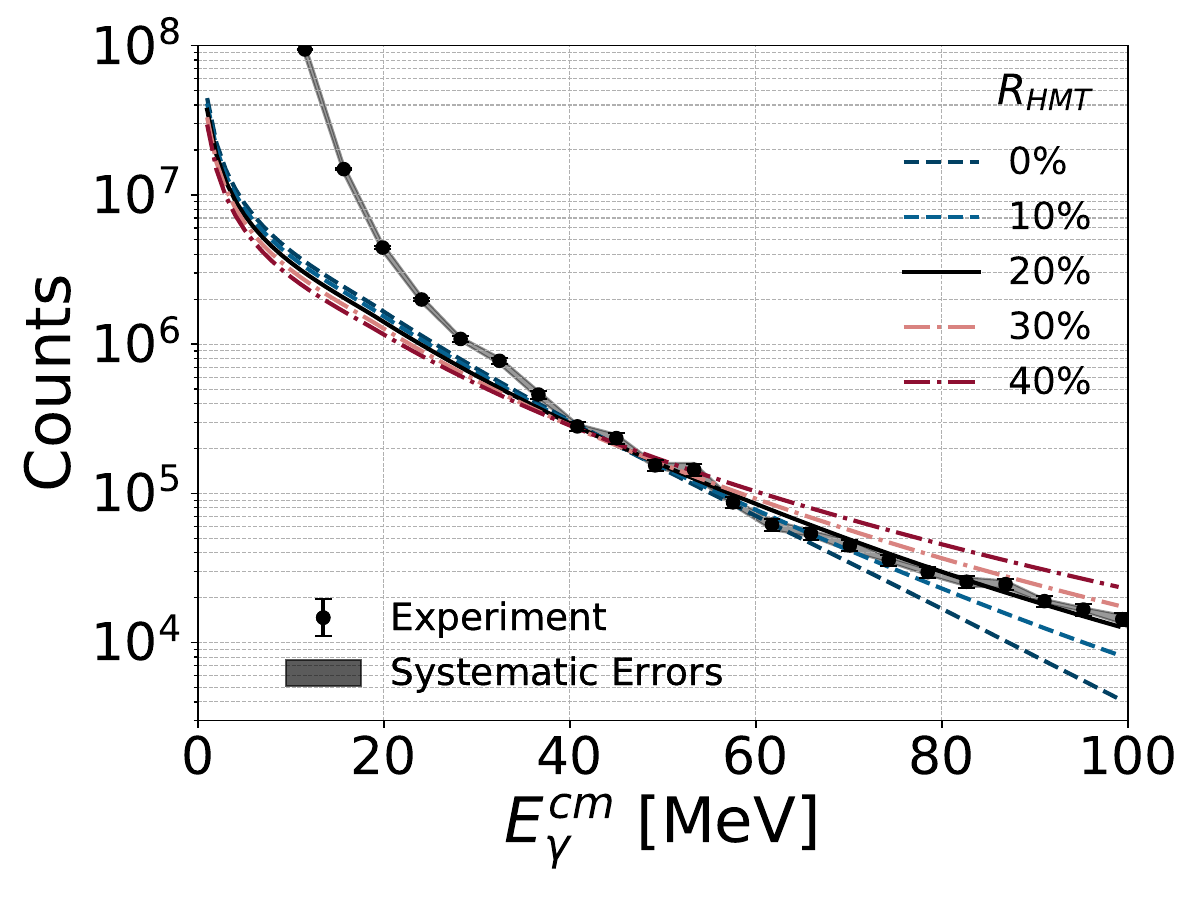}
    \caption{Comparison between several main theoretical curves and the reconstructed $\gamma$ spectrum.}
    \label{ReconstructedSpecwithErrors}
\end{figure}

After the iteration process described earlier, the reconstructed real $\gamma$ spectrum from the central measured spectrum is obtained and shown as black  dots in Fig.~\ref{ReconstructedSpecwithErrors}. 
The spectrum has been transformed from the lab frame to the center-of-mass frame and subsequently rebinned, combining every 4 bins into 1.

With the reconstructed real $\gamma$ spectrum, we can directly compare the experimental results with IBUU theoretical predictions without requiring any additional corrections or assumptions.

To present the comparison more clearly, the main theoretical curves are scaled by a specific coefficient of $4.5\times10^9$ and a rebin number to match the experimental data format. The comparison between the reconstructed spectrum and the scaled theoretical curves is shown in Fig.~\ref{ReconstructedSpecwithErrors}.

This direct comparison allows us to quantitatively evaluate the theoretical predictions and determine the HMT ratio that best describes the experimental data.

\subsection{Direct Comparison with IBUU-MDI Calculations}

To determine the favored ratio of HMT from the reconstructed spectrum using the Richardson-Lucy (RL) method, 
we define the $\chi^2(R_{\text{HMT}})$ to quantify the agreement between the experimentally measured spectrum and the theoretical predictions:
\begin{equation}
    \chi^2(R_{\text{HMT}})=\sum_i \frac{1}{\sigma_i^2}(\text{S}_i^{\text{exp}}-\alpha\text{S}_i^{R_{\text{HMT}}})^2,
\end{equation}
where $\text{S}_i^{\text{exp}}$ and $\text{S}_i^{R_{\text{HMT}}}$ represent the counts in the $i^{\rm th}$ energy bin for the experimental data and the theoretical prediction with a given $R_{\text{HMT}}$, respectively, and $\sigma_i$ is the uncertainty of the reconstructed experimental spectrum in the $i^{\rm th}$ bin. The parameter $\alpha$ is a normalization factor determined by minimizing $\chi^2$ for each theoretical curve.

Since the comparison focuses on the spectral shape rather than the absolute yield, the normalization factor only adjusts the overall amplitude of the theoretical spectrum. Therefore, this scaling does not influence the extracted value of $R_{\text{HMT}}$, which is determined by the differences in spectral shape.
The theoretical curve that minimizes $\chi^2(R_{\text{HMT}})$ corresponds to the best-fit $R_{\text{HMT}}$, providing the most accurate description of the experimental data.

\begin{figure}[hptb]
    \centering
    \includegraphics[width=0.95\linewidth]{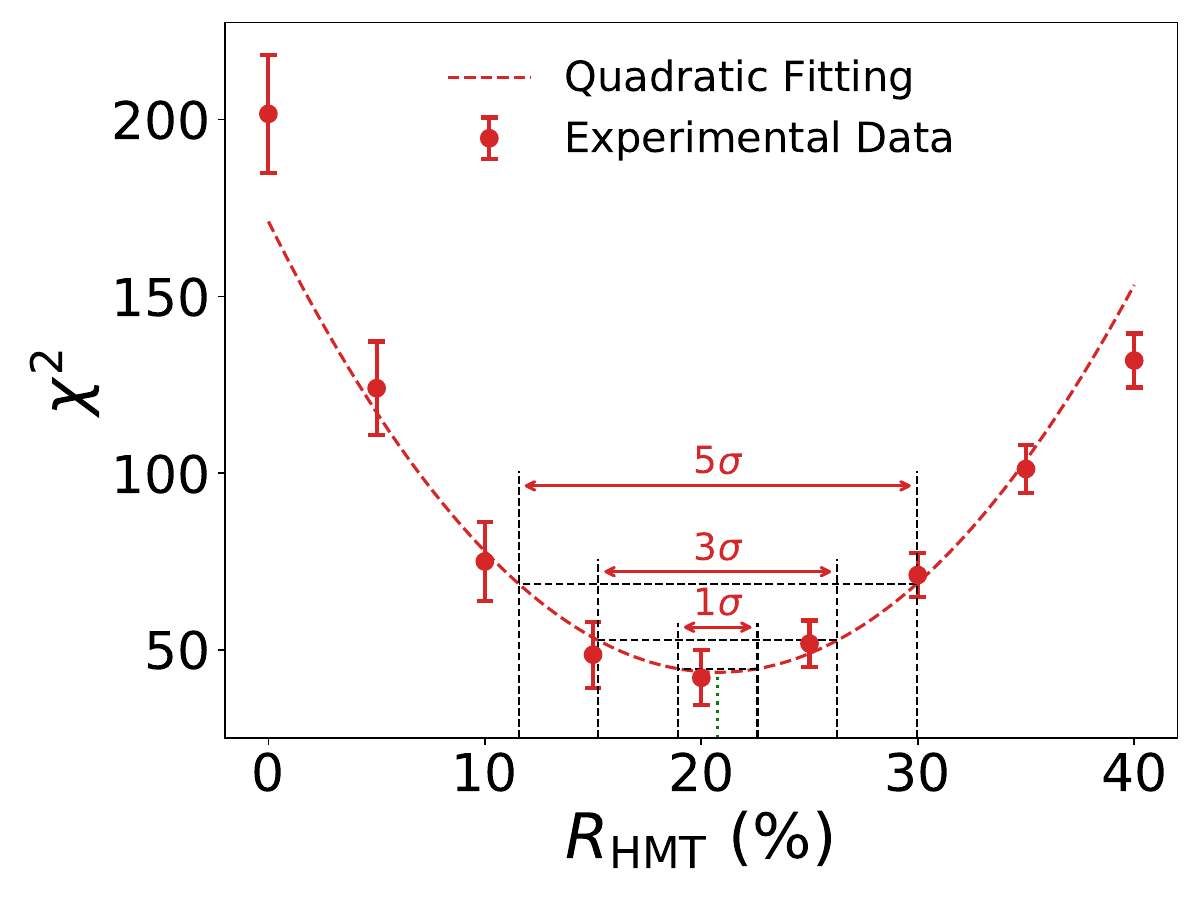}
    \caption{$\chi^2(R_{\text{HMT}})$ values for different $R_{\text{HMT}}$ values, along with the corresponding quadratic fit, in the energy range from $35 ~\rm MeV$  to $100 ~\rm MeV$.}
    \label{HMTComparisonsChi2}
\end{figure}

The energy range from $35~ \rm MeV$  to $100 ~\rm MeV$ is chosen as the central analysis interval, and the $\chi^2(R_{\text{HMT}})$ values are computed by comparing the reconstructed experimental spectrum with different theoretical curves. 
In the energy range of interest, the $\chi^2(R_{\text{HMT}})$ distribution is shown as red dots with error bars in Fig.~\ref{HMTComparisonsChi2}. A quadratic fit within the range of $5\% < R_{\text{HMT}} < 35\%$ to this distribution, represented by the dashed line, reveals that the minimum occurs at $R_{\text{HMT}} = (20.8\pm 0.2)\%$, with the uncertainty representing the fitting error.

The errors of each bin in the reconstructed $\gamma$ energy spectrum from experimental measurements by RL algorithm approximately follow a Gaussian distribution. Therefore, we can evaluate the confidence intervals for $R_{\text{HMT}}$ based on the $\chi^2$ distribution with one degree of freedom (one parameter $R_{\text{HMT}}$). Specifically, the values of $\chi^2$ that exceed the minimum by 1, 9 and 25 correspond to confidence levels of $\pm 1\sigma$, $\pm 3\sigma$ and $\pm 5\sigma$, respectively, as indicated by the black dashed lines in the figure. 
Therefore, the optimal value of $R_{\text{HMT}}$  determined from the reconstructed spectrum is 
\begin{equation}
    R_{\text{HMT}} = (20.8\pm 1.8)\%,
\end{equation}
This result is consistent with the forward analysis given by Eq. (\ref{rhmt_fnl}) within the error margin.

\section {Conclusion and Outlook} \label{conclusionoutlook}

In conclusion, the Bremsstrahlung $\gamma$-ray emissions were measured with high precision in $^{124}$Sn+$^{124}$Sn at 25 MeV/u using the experiment CSHINE.  In comparison to the previous experiment $^{86}$Kr+$^{124}$Sn, significant advancements were made in both the experimental setup and data analysis. These improvements include the installation of veto scintillators surrounding the CsI(Tl) hodoscope, the extension of the energy range compared to earlier experiments, and the testing of the non-linearity of the CsI(Tl) crystal response using quasi-monochromatic $\gamma$-rays from the Shanghai Laser Electron Gamma Source. The total statistics were also enhanced. Furthermore, the choice of symmetric reaction system enabled us to determine the HMT fraction ($R_{\rm HMT}$) for the specific nucleus $^{124}$Sn.

Two methods were employed to obtain the bremsstrahlung $\gamma$-ray energy spectrum. The primary method involved using the entire data set within a slow coincidence gate, with the cosmic-ray muon background subtracted, as measured with the beam off. The secondary method, used as a closure check, utilized data from a fast coincidence gate, with the background obtained from a beam-on random coincidence time window of the same width. The second method necessarily excluded the \ssdtwo~ trigger signal, as it is suffered from significant time jitter. Both approaches provided consistent constraints on the HMT fraction within the measurement uncertainties. From these analyses, we derived the fraction of high-momentum nucleons arising from SRC to be  $R_{\rm HMT}=(20 \pm 3)\%$.

From the experimentally measured spectrum, we also reconstructed the original bremsstrahlung $\gamma$-ray energy spectrum by applying the Richardson-Lucy deblurring algorithm, which originates from the application of the optical deblurring method to solve the inverse problem. The fraction of HMT nucleons inferred from this reconstruction was consistent with the result obtained above. This reconstruction of the original $\gamma$-ray spectrum enables direct comparisons with future theoretical calculations, independent of the detector response matrix.

The current experimental result provides a statistically precise determination of the SRC nucleon fraction in nuclei, confirming a non-zero $R_{\rm HMT}$ at a significance level greater than $5\sigma$. For the first time, our work establishes a new and complementary approach to probe SRC and parton dynamics in nuclei via low-energy heavy-ion collisions.

{\textbf{Acknowledgement~} } This work is supported by the National Natural Science Foundation of China under Grant Nos. 12335008, 12205160, 12275129, by the Ministry of Science and Technology under Grant No. 2022YFE0103400, by the Center for High Performance Computing and the Initiative Scientific Research Program in Tsinghua University and by HIRFL. The authors acknowledge the HIRFL machine team for providing excellent beam conditions. 

\bibliography{reference}

@article{Si:2024ujh,
    author = "Si, Dawei and others",
    title = "{The neutron array of the compact spectrometer for heavy ion experiments in Fermi energy region}",
    eprint = "2406.18605",
    archivePrefix = "arXiv",
    primaryClass = "physics.ins-det",
    doi = "10.1016/j.nima.2024.170055",
    journal = "Nucl. Instrum. Meth. A",
    volume = "1070",
    pages = "170055",
    year = "2025"
}

@article{Si:2025eou,
    author = "Si, Dawei and others",
    title = "{Extracting Neutron-Neutron Interaction Strength and Spatiotemporal Dynamics of Neutron Emission from the Two-Particle Correlation Function}",
    eprint = "2501.09576",
    archivePrefix = "arXiv",
    primaryClass = "nucl-ex",
    doi = "10.1103/PhysRevLett.134.222301",
    journal = "Phys. Rev. Lett.",
    volume = "134",
    number = "22",
    pages = "222301",
    year = "2025"
}

@article{Xu:2025mvv,
    author = "Xu, Junhuai and others",
    title = "{Precise measurement of short-range correlations in nuclei from bremsstrahlung gamma-ray emission in low-energy heavy-ion collisions}",
    eprint = "2504.13929",
    archivePrefix = "arXiv",
    primaryClass = "nucl-ex",
    doi = "10.1103/jw1p-36pb",
    journal = "Phys. Rev. Res.",
    volume = "7",
    number = "4",
    pages = "043174",
    year = "2025"
}

@article{XU2025170787,
	author = {Junhuai Xu and Dawei Si and Yuhao Qin and Mengke Xu and Kaijie Chen and Zirui Hao and Gongtao Fan and Hongwei Wang and Yijie Wang and Zhigang Xiao},
	doi = {https://doi.org/10.1016/j.nima.2025.170787},
	issn = {0168-9002},
	journal = {Nucl. Instrum. Meth. A},
	pages = {170787},
	title = {Linear response of CsI(Tl) crystal to energetic photons below 20 MeV},
	url = {https://www.sciencedirect.com/science/article/pii/S0168900225005881},
	volume = {1080},
	year = {2025}}

@article{Xu:2024oct,
    author = "Xu, JunHuai and Qin, Yuhao and Qin, Zhi and Si, Dawei and Zhang, Boyuan and Wang, Yijie and Niu, Qinglin and Xu, Chang and Xiao, Zhigang",
    title = "{Reconstruction of Bremsstrahlung \ensuremath{\gamma}-rays spectrum in heavy ion reactions with Richardson-Lucy algorithm}",
    eprint = "2405.06711",
    archivePrefix = "arXiv",
    primaryClass = "nucl-th",
    doi = "10.1016/j.physletb.2024.139009",
    journal = "Phys. Lett. B",
    volume = "857",
    pages = "139009",
    year = "2024"
}

@article{Qin:2023qcn,
    author = "Qin, Yuhao and others",
    title = "{Probing high-momentum component in nucleon momentum distribution by neutron-proton bremsstrahlung \ensuremath{\gamma}-rays in heavy ion reactions}",
    eprint = "2307.10717",
    archivePrefix = "arXiv",
    primaryClass = "nucl-ex",
    doi = "10.1016/j.physletb.2024.138514",
    journal = "Phys. Lett. B",
    volume = "850",
    pages = "138514",
    year = "2024"
}

@article{Qin:2022mzp,
    author = "Qin, Yuhao and others",
    title = "{A CsI(Tl) hodoscope on CSHINE for Bremsstrahlung \ensuremath{\gamma}-rays in heavy ion reactions}",
    eprint = "2212.13498",
    archivePrefix = "arXiv",
    primaryClass = "physics.ins-det",
    doi = "10.1016/j.nima.2023.168330",
    journal = "Nucl. Instrum. Meth. A",
    volume = "1053",
    pages = "168330",
    year = "2023"
}

@article{Wang:2021jgu,
    author = "Wang, Yi-Jie and others",
    title = "{CSHINE for studies of HBT correlation in Heavy Ion Reactions}",
    eprint = "2101.07352",
    archivePrefix = "arXiv",
    primaryClass = "physics.ins-det",
    doi = "10.1007/s41365-020-00842-2",
    journal = "Nucl. Sci. Tech.",
    volume = "32",
    number = "1",
    pages = "4",
    year = "2021"
}

@article{Wang:2021mrv,
    author = "Wang, Yijie and others",
    title = "{The emission order of hydrogen isotopes via correlation functions in 30 MeV/u Ar+Au reactions}",
    eprint = "2112.02210",
    archivePrefix = "arXiv",
    primaryClass = "nucl-ex",
    doi = "10.1016/j.physletb.2021.136856",
    journal = "Phys. Lett. B",
    volume = "825",
    pages = "136856",
    year = "2022"
}

@article{Wang:2022ysq,
    author = "Wang, Yijie and others",
    title = "{Observing the ping-pong modality of the isospin degree of freedom in cluster emission from heavy-ion reactions}",
    eprint = "2209.04079",
    archivePrefix = "arXiv",
    primaryClass = "nucl-ex",
    doi = "10.1103/PhysRevC.107.L041601",
    journal = "Phys. Rev. C",
    volume = "107",
    number = "4",
    pages = "L041601",
    year = "2023"
}

@article{Wang:2023kyp,
    author = "Wang, Yi-Jie and others",
    title = "{The enhancement of neutron-rich particle emission from out-of-fission-plane in Fermi energy heavy ion reactions}",
    eprint = "2311.07095",
    archivePrefix = "arXiv",
    primaryClass = "nucl-ex",
    doi = "10.1007/s41365-025-01742-z",
    journal = "Nucl. Sci. Tech.",
    volume = "36",
    number = "8",
    pages = "155",
    year = "2025"
}

@article{Guan:2021tbi,
    author = "Guan, Fenhai and others",
    title = "{A Compact Spectrometer for Heavy Ion Experiments in the Fermi energy regime}",
    doi = "10.1016/j.nima.2021.165592",
    journal = "Nucl. Instrum. Meth. A",
    volume = "1011",
    pages = "165592",
    year = "2021"
}

@article{Guan:2021nfk,
    author = "Guan, Fenhai and others",
    title = "{Track recognition for the \ensuremath{\Delta}E\ensuremath{-}E telescopes with silicon strip detectors}",
    eprint = "2110.08261",
    archivePrefix = "arXiv",
    primaryClass = "physics.ins-det",
    doi = "10.1016/j.nima.2022.166461",
    journal = "Nucl. Instrum. Meth. A",
    volume = "1029",
    pages = "166461",
    year = "2022"
}

@article{Guo:2022kwc,
    author = "Guo, Dong and others",
    title = "{An FPGA-based trigger system for CSHINE}",
    eprint = "2206.15382",
    archivePrefix = "arXiv",
    primaryClass = "physics.ins-det",
    doi = "10.1007/s41365-022-01149-0",
    journal = "Nucl. Sci. Tech.",
    volume = "33",
    number = "12",
    pages = "162",
    year = "2022"
}

@article{Wei:2025lbj,
    author = "Wei, Xiao-Bao and others",
    title = "{Data analysis framework for silicon strip detector in compact spectrometer for heavy-ion experiments}",
    doi = "10.1007/s41365-025-01743-y",
    journal = "Nucl. Sci. Tech.",
    volume = "36",
    number = "7",
    pages = "132",
    year = "2025"
}

@article{Zhang:2017xtk,
    author = "Zhang, Yan and others",
    title = "{Long-time drift of the isospin degree of freedom in heavy ion collisions}",
    doi = "10.1103/PhysRevC.95.041602",
    journal = "Phys. Rev. C",
    volume = "95",
    number = "4",
    pages = "041602",
    year = "2017"
}

@article{Frankfurt:1988nt,
    author = "Frankfurt, L. L. and Strikman, M. I.",
    title = "{Hard Nuclear Processes and Microscopic Nuclear Structure}",
    doi = "10.1016/0370-1573(88)90179-2",
    journal = "Phys. Rept.",
    volume = "160",
    pages = "235--427",
    year = "1988"
}

@article{CiofiDegliAtti:1989eg,
    author = "Ciofi Degli Atti, Claudio and Liuti, S.",
    title = "{On the Effects of Nucleon Binding and Correlations in Deep Inelastic Electron Scattering by Nuclei}",
    doi = "10.1016/0370-2693(89)90808-3",
    journal = "Phys. Lett. B",
    volume = "225",
    pages = "215--221",
    year = "1989"
}

@article{Hagel.WPCF2023,
title = {Possible signatures of short range correlations in intermediate energy heavy ion collisions},
author = "{K. Hagel}",
Journal = {IL Nuovo Cimento},
volume = {48C},
pages = {30},
year = {2025},
doi = {10.1393/ncc/i2025-25030-5},
url = {https://www.sif.it/riviste/sif/ncc/econtents/2025/048/01/article/29}
}

@article{Huang:2025uvc,
    author = "Huang, Yu-Jing and Meng, Zhu and Pang, Long-Gang and Wang, Xin-Nian",
    title = "{A Novel Deep Learning Method for Detecting Nucleon-Nucleon Correlations}",
    eprint = "2504.00790",
    archivePrefix = "arXiv",
    primaryClass = "nucl-th",
    year = "2025",
    month = "4",
    journal = "arXiv preprint"
}

@article{Arrington:2011xs,
    author = "Arrington, J. and Higinbotham, D. W. and Rosner, G. and Sargsian, M.",
    title = "{Hard probes of short-range nucleon-nucleon correlations}",
    eprint = "1104.1196",
    archivePrefix = "arXiv",
    primaryClass = "nucl-ex",
    reportNumber = "PHY-12946-ME-2011, JLAB-PHY-11-1329",
    doi = "10.1016/j.ppnp.2012.04.002",
    journal = "Prog. Part. Nucl. Phys.",
    volume = "67",
    pages = "898--938",
    year = "2012"
}

@article{nCTEQ:2023cpo,
    author = "Denniston, A. W. and others",
    collaboration = "nCTEQ",
    title = "{Modification of Quark-Gluon Distributions in Nuclei by Correlated Nucleon Pairs}",
    eprint = "2312.16293",
    archivePrefix = "arXiv",
    primaryClass = "hep-ph",
    reportNumber = "MS-TP-22-13, IFJPAN-IV-2022-21, SMU-HEP-23-08, FERMILAB-PUB-23-0840, SMU-HEP-22-13, FNAL-PUB-23-146-ND",
    doi = "10.1103/PhysRevLett.133.152502",
    journal = "Phys. Rev. Lett.",
    volume = "133",
    number = "15",
    pages = "152502",
    year = "2024"
}

@article{EuropeanMuon:1983wih,
    author = "Aubert, J. J. and others",
    collaboration = "European Muon",
    title = "{The ratio of the nucleon structure functions $F2_n$ for iron and deuterium}",
    reportNumber = "CERN-EP/83-14",
    doi = "10.1016/0370-2693(83)90437-9",
    journal = "Phys. Lett. B",
    volume = "123",
    pages = "275--278",
    year = "1983"
}

@article{Geesaman:1995yd,
    author = "Geesaman, Donald F. and Saito, K. and Thomas, Anthony William",
    title = "{The nuclear EMC effect}",
    reportNumber = "PHY-7963-ME-95",
    doi = "10.1146/annurev.ns.45.120195.002005",
    journal = "Ann. Rev. Nucl. Part. Sci.",
    volume = "45",
    pages = "337--390",
    year = "1995"
}

@article{Kelly:1996hd,
    author = "Kelly, J. J.",
    editor = "Negele, John W. and Vogt, E.",
    title = "{Nucleon knockout by intermediate-energy electrons}",
    doi = "10.1007/0-306-47067-5_2",
    journal = "Adv. Nucl. Phys.",
    volume = "23",
    pages = "75--294",
    year = "1996"
}

@article{Norton:2003cb,
    author = "Norton, P. R.",
    title = "{The EMC effect}",
    doi = "10.1088/0034-4885/66/8/201",
    journal = "Rept. Prog. Phys.",
    volume = "66",
    pages = "1253--1297",
    year = "2003"
}

@article{Dickhoff:2004xx,
    author = "Dickhoff, W. H. and Barbieri, C.",
    title = "{Selfconsistent Green's function method for nuclei and nuclear matter}",
    eprint = "nucl-th/0402034",
    archivePrefix = "arXiv",
    reportNumber = "TRI-PP-03-40",
    doi = "10.1016/j.ppnp.2004.02.038",
    journal = "Prog. Part. Nucl. Phys.",
    volume = "52",
    pages = "377--496",
    year = "2004"
}

@article{Malace:2014uea,
    author = "Malace, Simona and Gaskell, David and Higinbotham, Douglas W. and Cloet, Ian",
    title = "{The Challenge of the EMC Effect: existing data and future directions}",
    eprint = "1405.1270",
    archivePrefix = "arXiv",
    primaryClass = "nucl-ex",
    reportNumber = "JLAB-PHY-14-1886",
    doi = "10.1142/S0218301314300136",
    journal = "Int. J. Mod. Phys. E",
    volume = "23",
    number = "08",
    pages = "1430013",
    year = "2014"
}

@article{Kovarik:2015cma,
    author = "Kovarik, K. and others",
    title = "{nCTEQ15 - Global analysis of nuclear parton distributions with uncertainties in the CTEQ framework}",
    eprint = "1509.00792",
    archivePrefix = "arXiv",
    primaryClass = "hep-ph",
    reportNumber = "LPSC-15-153, MS-TP-15-11, FERMILAB-PUB-15-375-ND-PPD-T",
    doi = "10.1103/PhysRevD.93.085037",
    journal = "Phys. Rev. D",
    volume = "93",
    number = "8",
    pages = "085037",
    year = "2016"
}

@article{Eskola:2021nhw,
    author = "Eskola, Kari J. and Paakkinen, Petja and Paukkunen, Hannu and Salgado, Carlos A.",
    title = "{EPPS21: a global QCD analysis of nuclear PDFs}",
    eprint = "2112.12462",
    archivePrefix = "arXiv",
    primaryClass = "hep-ph",
    doi = "10.1140/epjc/s10052-022-10359-0",
    journal = "Eur. Phys. J. C",
    volume = "82",
    number = "5",
    pages = "413",
    year = "2022"
}

@article{Helenius:2021tof,
    author = "Helenius, Ilkka and Walt, Marina and Vogelsang, Werner",
    title = "{NNLO nuclear parton distribution functions with electroweak-boson production data from the LHC}",
    eprint = "2112.11904",
    archivePrefix = "arXiv",
    primaryClass = "hep-ph",
    doi = "10.1103/PhysRevD.105.094031",
    journal = "Phys. Rev. D",
    volume = "105",
    number = "9",
    pages = "094031",
    year = "2022"
}

@article{AbdulKhalek:2022fyi,
    author = "Abdul Khalek, Rabah and Gauld, Rhorry and Giani, Tommaso and Nocera, Emanuele R. and Rabemananjara, Tanjona R. and Rojo, Juan",
    title = "{nNNPDF3.0: evidence for a modified partonic structure in heavy nuclei}",
    eprint = "2201.12363",
    archivePrefix = "arXiv",
    primaryClass = "hep-ph",
    reportNumber = "Nikhef-2021-028, BONN-TH-2021-14",
    doi = "10.1140/epjc/s10052-022-10417-7",
    journal = "Eur. Phys. J. C",
    volume = "82",
    number = "6",
    pages = "507",
    year = "2022"
}

@article{Segarra:2020gtj,
    author = "Segarra, E. P. and others",
    title = "{Extending nuclear PDF analyses into the high-$x$ , low-$Q^2$ region}",
    eprint = "2012.11566",
    archivePrefix = "arXiv",
    primaryClass = "hep-ph",
    reportNumber = "FERMILAB-PUB-20-668-E, FNAL-PUB-20-668, JLAB-THY-20-3303, SMU-HEP-20-07, MS-TP-20-40, KA-TP-16-2020, MS-TP-20-40,
  KA-TP-16-2020, P3H-20-052, IFJPAN-IV-2020-9",
    doi = "10.1103/PhysRevD.103.114015",
    journal = "Phys. Rev. D",
    volume = "103",
    number = "11",
    pages = "114015",
    year = "2021"
}

@article{Ruiz:2023ozv,
    author = "Ruiz, R. and others",
    title = "{Target mass corrections in lepton\textendash{}nucleus DIS: Theory and applications to nuclear PDFs}",
    eprint = "2301.07715",
    archivePrefix = "arXiv",
    primaryClass = "hep-ph",
    reportNumber = "IFJPAN-IV-2022-18, SMU-HEP-22-12, MS-TP-22-49, ANL-180568, FERMILAB-PUB-23-022-ND, JLAB-THY-23-3748",
    doi = "10.1016/j.ppnp.2023.104096",
    journal = "Prog. Part. Nucl. Phys.",
    volume = "136",
    pages = "104096",
    year = "2024"
}

@article{Piasetzky:2006ai,
    author = "Piasetzky, E. and Sargsian, M. and Frankfurt, L. and Strikman, M. and Watson, J. W.",
    title = "{Evidence for the strong dominance of proton-neutron correlations in nuclei}",
    eprint = "nucl-th/0604012",
    archivePrefix = "arXiv",
    reportNumber = "FIU-NUPAR-04-06",
    doi = "10.1103/PhysRevLett.97.162504",
    journal = "Phys. Rev. Lett.",
    volume = "97",
    pages = "162504",
    year = "2006"
}

@article{Subedi:2008zz,
    author = "Subedi, R. and others",
    title = "{Probing Cold Dense Nuclear Matter}",
    eprint = "0908.1514",
    archivePrefix = "arXiv",
    primaryClass = "nucl-ex",
    reportNumber = "JLAB-PHY-08-828",
    doi = "10.1126/science.1156675",
    journal = "Science",
    volume = "320",
    pages = "1476--1478",
    year = "2008"
}

@article{LabHallA:2014wqo,
    author = "Korover, I. and others",
    collaboration = "Lab Hall A",
    title = "{Probing the Repulsive Core of the Nucleon-Nucleon Interaction via the $^4He(e,e^{\prime}pN)$ Triple-Coincidence Reaction}",
    eprint = "1401.6138",
    archivePrefix = "arXiv",
    primaryClass = "nucl-ex",
    reportNumber = "JLAB-PHY-14-1862",
    doi = "10.1103/PhysRevLett.113.022501",
    journal = "Phys. Rev. Lett.",
    volume = "113",
    number = "2",
    pages = "022501",
    year = "2014"
}

@article{Hen:2014nza,
    author = "Hen, O. and others",
    title = "{Momentum sharing in imbalanced Fermi systems}",
    eprint = "1412.0138",
    archivePrefix = "arXiv",
    primaryClass = "nucl-ex",
    doi = "10.1126/science.1256785",
    journal = "Science",
    volume = "346",
    pages = "614--617",
    year = "2014"
}

@article{CLAS:2018xvc,
    author = "Duer, M. and others",
    collaboration = "CLAS",
    title = "{Direct Observation of Proton-Neutron Short-Range Correlation Dominance in Heavy Nuclei}",
    eprint = "1810.05343",
    archivePrefix = "arXiv",
    primaryClass = "nucl-ex",
    doi = "10.1103/PhysRevLett.122.172502",
    journal = "Phys. Rev. Lett.",
    volume = "122",
    number = "17",
    pages = "172502",
    year = "2019"
}

@article{CLAS:2020rue,
    author = "Korover, I. and others",
    collaboration = "CLAS",
    title = "{12C(e,e'pN) measurements of short range correlations in the tensor-to-scalar interaction transition region}",
    eprint = "2004.07304",
    archivePrefix = "arXiv",
    primaryClass = "nucl-ex",
    doi = "10.1016/j.physletb.2021.136523",
    journal = "Phys. Lett. B",
    volume = "820",
    pages = "136523",
    year = "2021"
}

@article{West:2020tyo,
    author = "West, Jennifer Rittenhouse",
    title = "{Diquark induced short-range nucleon-nucleon correlations \& the EMC effect}",
    eprint = "2009.06968",
    archivePrefix = "arXiv",
    primaryClass = "hep-ph",
    doi = "10.1016/j.nuclphysa.2022.122563",
    journal = "Nucl. Phys. A",
    volume = "1029",
    pages = "122563",
    year = "2023"
}

@article{Schiavilla:2006xx,
    author = "Schiavilla, R. and Wiringa, Robert B. and Pieper, Steven C. and Carlson, J.",
    title = "{Tensor Forces and the Ground-State Structure of Nuclei}",
    eprint = "nucl-th/0611037",
    archivePrefix = "arXiv",
    reportNumber = "JLAB-THY-06-562",
    doi = "10.1103/PhysRevLett.98.132501",
    journal = "Phys. Rev. Lett.",
    volume = "98",
    pages = "132501",
    year = "2007"
}

@article{Wiringa:2008dn,
    author = "Wiringa, R. B. and Schiavilla, R. and Pieper, Steven C. and Carlson, J.",
    title = "{Dependence of two-nucleon momentum densities on total pair momentum}",
    eprint = "0806.1718",
    archivePrefix = "arXiv",
    primaryClass = "nucl-th",
    reportNumber = "JLAB-THY-08-826",
    doi = "10.1103/PhysRevC.78.021001",
    journal = "Phys. Rev. C",
    volume = "78",
    pages = "021001",
    year = "2008"
}

@article{Wiringa:2013ala,
    author = "Wiringa, R. B. and Schiavilla, R. and Pieper, Steven C. and Carlson, J.",
    title = "{Nucleon and nucleon-pair momentum distributions in $A \le 12$ nuclei}",
    eprint = "1309.3794",
    archivePrefix = "arXiv",
    primaryClass = "nucl-th",
    reportNumber = "JLAB-THY-13-1789",
    doi = "10.1103/PhysRevC.89.024305",
    journal = "Phys. Rev. C",
    volume = "89",
    number = "2",
    pages = "024305",
    year = "2014"
}

@article{CLAS:2018yvt,
    author = "Duer, M. and others",
    collaboration = "CLAS",
    title = "{Probing high-momentum protons and neutrons in neutron-rich nuclei}",
    doi = "10.1038/s41586-018-0400-z",
    journal = "Nature",
    volume = "560",
    number = "7720",
    pages = "617--621",
    year = "2018"
}

@article{Wei:2019jva,
    author = "Wei, Gao-Feng and Cao, Xi-Guang and Zhi, Qi-Jun and Cao, Xin-Wei and Long, Zheng-Wen",
    title = "{Proton-proton momentum correlation function as a probe of the high momentum tail of the nucleon momentum distribution}",
    eprint = "1912.03165",
    archivePrefix = "arXiv",
    primaryClass = "nucl-th",
    doi = "10.1103/PhysRevC.101.014613",
    journal = "Phys. Rev. C",
    volume = "101",
    number = "1",
    pages = "014613",
    year = "2020"
}

@article{Hen:2016kwk,
    author = "Hen, O. and Miller, G. A. and Piasetzky, E. and Weinstein, L. B.",
    title = "{Nucleon-Nucleon Correlations, Short-lived Excitations, and the Quarks Within}",
    eprint = "1611.09748",
    archivePrefix = "arXiv",
    primaryClass = "nucl-ex",
    doi = "10.1103/RevModPhys.89.045002",
    journal = "Rev. Mod. Phys.",
    volume = "89",
    number = "4",
    pages = "045002",
    year = "2017"
}

@article{JeffersonLabHallA:2007lly,
    author = "Shneor, R. and others",
    collaboration = "Jefferson Lab Hall A",
    title = "{Investigation of proton-proton short-range correlations via the C-12(e, e-prime pp) reaction}",
    eprint = "nucl-ex/0703023",
    archivePrefix = "arXiv",
    reportNumber = "JLAB-PHY-07-624",
    doi = "10.1103/PhysRevLett.99.072501",
    journal = "Phys. Rev. Lett.",
    volume = "99",
    pages = "072501",
    year = "2007"
}

@article{CLAS:2010yvl,
    author = "Baghdasaryan, H. and others",
    collaboration = "CLAS",
    title = "{Tensor Correlations Measured in $^3He(e,e'pp)n$}",
    eprint = "1008.3100",
    archivePrefix = "arXiv",
    primaryClass = "nucl-ex",
    reportNumber = "JLAB-PHY-10-1237",
    doi = "10.1103/PhysRevLett.105.222501",
    journal = "Phys. Rev. Lett.",
    volume = "105",
    pages = "222501",
    year = "2010"
}

@article{Frankfurt:1993sp,
    author = "Frankfurt, L. L. and Strikman, M. I. and Day, D. B. and Sargsian, M.",
    title = "{Evidence for short range correlations from high Q**2 (e, e-prime) reactions}",
    doi = "10.1103/PhysRevC.48.2451",
    journal = "Phys. Rev. C",
    volume = "48",
    pages = "2451--2461",
    year = "1993"
}

@article{Arrington:1998ps,
    author = "Arrington, J. and others",
    title = "{Inclusive electron - nucleus scattering at large momentum transfer}",
    eprint = "nucl-ex/9811008",
    archivePrefix = "arXiv",
    reportNumber = "OAP-747",
    doi = "10.1103/PhysRevLett.82.2056",
    journal = "Phys. Rev. Lett.",
    volume = "82",
    pages = "2056--2059",
    year = "1999"
}

@article{CLAS:2003eih,
    author = "Egiyan, K. S. and others",
    collaboration = "CLAS",
    title = "{Observation of nuclear scaling in the A(e, e-prime) reaction at x(B) greater than 1}",
    eprint = "nucl-ex/0301008",
    archivePrefix = "arXiv",
    reportNumber = "JLAB-PHY-03-48",
    doi = "10.1103/PhysRevC.68.014313",
    journal = "Phys. Rev. C",
    volume = "68",
    pages = "014313",
    year = "2003"
}

@article{CLAS:2005ola,
    author = "Egiyan, K. S. and others",
    collaboration = "CLAS",
    title = "{Measurement of 2- and 3-nucleon short range correlation probabilities in nuclei}",
    eprint = "nucl-ex/0508026",
    archivePrefix = "arXiv",
    reportNumber = "JLAB-PHY-05-285",
    doi = "10.1103/PhysRevLett.96.082501",
    journal = "Phys. Rev. Lett.",
    volume = "96",
    pages = "082501",
    year = "2006"
}

@article{Fomin:2011ng,
    author = "Fomin, N. and others",
    title = "{New measurements of high-momentum nucleons and short-range structures in nuclei}",
    eprint = "1107.3583",
    archivePrefix = "arXiv",
    primaryClass = "nucl-ex",
    reportNumber = "JLAB-PHY-11-1408",
    doi = "10.1103/PhysRevLett.108.092502",
    journal = "Phys. Rev. Lett.",
    volume = "108",
    pages = "092502",
    year = "2012"
}

@article{CLAS:2019vsb,
    author = "Schmookler, B. and others",
    collaboration = "CLAS",
    title = "{Modified structure of protons and neutrons in correlated pairs}",
    eprint = "2004.12065",
    archivePrefix = "arXiv",
    primaryClass = "nucl-ex",
    doi = "10.1038/s41586-019-0925-9",
    journal = "Nature",
    volume = "566",
    number = "7744",
    pages = "354--358",
    year = "2019"
}

@article{Hen:2012fm,
    author = "Hen, O. and Piasetzky, E. and Weinstein, L. B.",
    title = "{New data strengthen the connection between Short Range Correlations and the EMC effect}",
    eprint = "1202.3452",
    archivePrefix = "arXiv",
    primaryClass = "nucl-ex",
    doi = "10.1103/PhysRevC.85.047301",
    journal = "Phys. Rev. C",
    volume = "85",
    pages = "047301",
    year = "2012"
}

@article{Tang:2002ww,
    author = "Tang, A. and others",
    title = "{n-p short range correlations from (p,2p + n) measurements}",
    eprint = "nucl-ex/0206003",
    archivePrefix = "arXiv",
    reportNumber = "KSU-CNR-202-07",
    doi = "10.1103/PhysRevLett.90.042301",
    journal = "Phys. Rev. Lett.",
    volume = "90",
    pages = "042301",
    year = "2003"
}

@article{Wakasa:2017rsk,
    author = "Wakasa, T. and Ogata, K. and Noro, T.",
    title = "{Proton-induced knockout reactions with polarized and unpolarized beams}",
    doi = "10.1016/j.ppnp.2017.06.002",
    journal = "Prog. Part. Nucl. Phys.",
    volume = "96",
    pages = "32--87",
    year = "2017"
}

@article{CiofidegliAtti:2015lcu,
    author = "Ciofi degli Atti, Claudio",
    title = "{In-medium short-range dynamics of nucleons: Recent theoretical and experimental advances}",
    doi = "10.1016/j.physrep.2015.06.002",
    journal = "Phys. Rept.",
    volume = "590",
    pages = "1--85",
    year = "2015"
}

@article{Patsyuk:2021fju,
    author = "Patsyuk, M. and others",
    title = "{Unperturbed inverse kinematics nucleon knockout measurements with a 48 GeV/c carbon beam}",
    eprint = "2102.02626",
    archivePrefix = "arXiv",
    primaryClass = "nucl-ex",
    doi = "10.1038/s41567-021-01193-4",
    journal = "Nature Phys.",
    volume = "17",
    pages = "693",
    year = "2021"
}

@article{Njock:1986uzx,
    author = "Njock, M. Kwato and Maurel, M. and Monnand, E. and Nifenecker, H. and Pinston, J. and Schussler, F. and Barneoud, D.",
    title = "{High energy gamma-ray production in heavy-ion reactions}",
    doi = "10.1016/0370-2693(86)90700-8",
    journal = "Phys. Lett. B",
    volume = "175",
    pages = "125--128",
    year = "1986"
}

@article{Noll:1984fd,
    author = "Noll, H. and others",
    title = "{Cooperative Effects Observed in the $\pi^0$ Production From Nucleus-nucleus Collisions}",
    reportNumber = "CERN-EP/84-32",
    doi = "10.1103/PhysRevLett.52.1284",
    journal = "Phys. Rev. Lett.",
    volume = "52",
    pages = "1284",
    year = "1984"
}

@article{Stevenson:1986zz,
    author = "Stevenson, J. and others",
    title = "{High-Energy Gamma-Ray Emission in Heavy-Ion Collisions}",
    doi = "10.1103/PhysRevLett.57.555",
    journal = "Phys. Rev. Lett.",
    volume = "57",
    pages = "555--558",
    year = "1986"
}

@article{Bauer:1986zz,
    author = "Bauer, W. and Bertsch, G. F. and Cassing, Wolfgang and Mosel, Ulrich",
    title = "{Energetic photons from intermediate energy proton- and heavy-ion-induced reactions}",
    doi = "10.1103/PhysRevC.34.2127",
    journal = "Phys. Rev. C",
    volume = "34",
    pages = "2127--2133",
    year = "1986"
}

@article{PhysRevC.53.R553,
  title = {Analysis of hard two-photon correlations measured in heavy-ion reactions at intermediate energies},
  author = {Barz, H. W. and K\"ampfer, B. and Wolf, Gy. and Bauer, W.},
  journal = {Phys. Rev. C},
  volume = {53},
  issue = {2},
  pages = {R553--R557},
  numpages = {0},
  year = {1996},
  month = {Feb},
  publisher = {American Physical Society},
  doi = {10.1103/PhysRevC.53.R553},
  url = {https://link.aps.org/doi/10.1103/PhysRevC.53.R553}
}

@article{Xue:2016udl,
    author = "Xue, Hui and Xu, Chang and Yong, Gao-Chan and Ren, Zhongzhou",
    title = "{Neutron\textendash{}proton bremsstrahlung as a possible probe of high-momentum component in nucleon momentum distribution}",
    doi = "10.1016/j.physletb.2016.02.044",
    journal = "Phys. Lett. B",
    volume = "755",
    pages = "486--490",
    year = "2016"
}

@article{Guo:2021zcs,
    author = "Guo, Wen-Mei and Li, Bao-An and Yong, Gao-Chan",
    title = "{Imprints of high-momentum nucleons in nuclei on hard photons from heavy-ion collisions near the Fermi energy}",
    eprint = "2106.08242",
    archivePrefix = "arXiv",
    primaryClass = "nucl-th",
    doi = "10.1103/PhysRevC.104.034603",
    journal = "Phys. Rev. C",
    volume = "104",
    number = "3",
    pages = "034603",
    year = "2021"
}

@article{Xu:2012hf,
    author = "Xu, Chang and Li, Ang and Li, Bao-An",
    editor = "Li, Bao-An and Natowitz, Joseph",
    title = "{Delineating effects of tensor force on the density dependence of nuclear symmetry energy}",
    eprint = "1207.1639",
    archivePrefix = "arXiv",
    primaryClass = "nucl-th",
    doi = "10.1088/1742-6596/420/1/012090",
    journal = "J. Phys. Conf. Ser.",
    volume = "420",
    pages = "012090",
    year = "2013"
}

@article{Remington:1987zza,
    author = "Remington, B. A. and Blann, M. and Bertsch, G. F.",
    title = "{n-p bremsstrahlung interpretation of high energy gamma rays from heavy-ion collisions}",
    doi = "10.1103/PhysRevC.35.1720",
    journal = "Phys. Rev. C",
    volume = "35",
    pages = "1720--1729",
    year = "1987"
}

@article{Ko:1985hq,
    author = "Ko, C. M. and Bertsch, G. and Aichelin, J.",
    title = "{PROBING HEAVY ION COLLISIONS WITH BREMSSTRAHLUNG}",
    doi = "10.1103/PhysRevC.31.2324",
    journal = "Phys. Rev. C",
    volume = "31",
    pages = "2324--2326",
    year = "1985"
}

@article{Jetter:1994bc,
    author = "Jetter, M. and von Geramb, H. V.",
    title = "{Nucleon-nucleon potentials and their test with bremsstrahlung}",
    doi = "10.1103/PhysRevC.49.1832",
    journal = "Phys. Rev. C",
    volume = "49",
    pages = "1832--1836",
    year = "1994"
}

@article{Wang:2020bzn,
    author = "Wang, S. S. and Ma, Y. G. and Cao, X. G. and Fang, D. Q. and Ma, C. W.",
    title = "{Hard-photon production and its correlation with intermediate-mass fragments in a framework of a quantum molecular dynamics model}",
    doi = "10.1103/PhysRevC.102.024620",
    journal = "Phys. Rev. C",
    volume = "102",
    number = "2",
    pages = "024620",
    year = "2020"
}

@article{vanGoethem:2001hy,
    author = "van Goethem, M. J. and others",
    title = "{Suppression of soft nuclear bremsstrahlung in proton - nucleus collisions}",
    eprint = "nucl-ex/0111021",
    archivePrefix = "arXiv",
    doi = "10.1103/PhysRevLett.88.122302",
    journal = "Phys. Rev. Lett.",
    volume = "88",
    pages = "122302",
    year = "2002"
}

@article{Maydanyuk:2013aqa,
    author = "Maydanyuk, Sergei P. and Zhang, Peng-Ming",
    title = "{New approach to determine proton-nucleus interactions from experimental bremsstrahlung data}",
    eprint = "1309.2784",
    archivePrefix = "arXiv",
    primaryClass = "nucl-th",
    doi = "10.1103/PhysRevC.91.024605",
    journal = "Phys. Rev. C",
    volume = "91",
    number = "2",
    pages = "024605",
    year = "2015"
}

@article{Li:2018lpy,
    author = "Li, Bao-An and Cai, Bao-Jun and Chen, Lie-Wen and Xu, Jun",
    title = "{Nucleon Effective Masses in Neutron-Rich Matter}",
    eprint = "1801.01213",
    archivePrefix = "arXiv",
    primaryClass = "nucl-th",
    doi = "10.1016/j.ppnp.2018.01.001",
    journal = "Prog. Part. Nucl. Phys.",
    volume = "99",
    pages = "29--119",
    year = "2018"
}

@article{Das:2002fr,
    author = "Das, C. B. and Gupta, S. Das and Gale, Charles and Li, Bao-An",
    title = "{Momentum dependence of symmetry potential in asymmetric nuclear matter for transport model calculations}",
    eprint = "nucl-th/0212090",
    archivePrefix = "arXiv",
    doi = "10.1103/PhysRevC.67.034611",
    journal = "Phys. Rev. C",
    volume = "67",
    pages = "034611",
    year = "2003"
}

@article{Mamba:2024pch,
    author = "Mamba, Sinethemba Neliswa and Danielewicz, Pawel",
    title = "{Analysis of Iterative Deblurring: No Explicit Noise}",
    eprint = "2407.03458",
    archivePrefix = "arXiv",
    primaryClass = "math.NA",
    month = "7",
    year = "2024",
    journal = "arXiv preprint"
}

@article{Xu:2024dnd,
    author = "Xu, Junhuai and Qin, Zhi and Zou, Renjie and Si, Dawei and Xiao, Sheng and Tian, Baiting and Wang, Yijie and Xiao, Zhigang",
    title = "{Imaging Freeze-Out Sources and Extracting Strong Interaction Parameters in Relativistic Heavy-Ion Collisions}",
    eprint = "2411.08718",
    archivePrefix = "arXiv",
    primaryClass = "nucl-th",
    doi = "10.1088/0256-307X/42/3/031401",
    journal = "Chin. Phys. Lett.",
    volume = "42",
    number = "3",
    pages = "031401",
    year = "2025"
}

@article{Nzabahimana:2025qdj,
    author = "Nzabahimana, Pierre and Lovell, Amy E. and Talou, Patrick",
    title = "{Deblurring fission fragment mass distributions}",
    eprint = "2505.01294",
    archivePrefix = "arXiv",
    primaryClass = "nucl-th",
    doi = "10.1103/dk6w-xzy8",
    journal = "Phys. Rev. C",
    volume = "112",
    number = "6",
    pages = "064607",
    year = "2025"
}

@article{Nzabahimana:2023tab,
    author = "Nzabahimana, Pierre and Danielewicz, Pawel",
    title = "{Source function from two-particle correlation through deblurring}",
    eprint = "2307.00173",
    archivePrefix = "arXiv",
    primaryClass = "nucl-th",
    doi = "10.1016/j.physletb.2023.138247",
    journal = "Phys. Lett. B",
    volume = "846",
    pages = "138247",
    year = "2023"
}

@article{Danielewicz:2021vqq,
    author = "Danielewicz, Pawel and Kurata-Nishimura, Mizuki",
    title = "{Deblurring for nuclei: 3D characteristics of heavy-ion collisions}",
    eprint = "2109.02626",
    archivePrefix = "arXiv",
    primaryClass = "nucl-th",
    doi = "10.1103/PhysRevC.105.034608",
    journal = "Phys. Rev. C",
    volume = "105",
    number = "3",
    pages = "034608",
    year = "2022"
}

@article{Nzabahimana:2022ndq,
    author = "Nzabahimana, Pierre and Redpath, Thomas and Baumann, Thomas and Danielewicz, Pawel and Giuliani, Pablo and Gu\`eye, Paul",
    title = "{Deconvoluting experimental decay energy spectra: The O26 case}",
    eprint = "2210.00157",
    archivePrefix = "arXiv",
    primaryClass = "nucl-th",
    doi = "10.1103/PhysRevC.107.064315",
    journal = "Phys. Rev. C",
    volume = "107",
    number = "6",
    pages = "064315",
    year = "2023"
}

@article{Vargas:2013kga,
    author = "Vargas, J. and Benlliure, J. and Caamano, M.",
    title = "{Unfolding the response of a zero-degree magnetic spectrometer from measurements of the $\Delta$ resonance}",
    doi = "10.1016/j.nima.2012.12.087",
    journal = "Nucl. Instrum. Meth. A",
    volume = "707",
    pages = "16--25",
    year = "2013"
}

@article{Li:1996ix,
    author = "Li, Bao-An and Ren, Zhong-Zhou and Ko, C. M. and Yennello, Sherry J.",
    title = "{Isospin dependence of collective flow in heavy ion collisions at intermediate-energies}",
    eprint = "nucl-th/9605015",
    archivePrefix = "arXiv",
    doi = "10.1103/PhysRevLett.76.4492",
    journal = "Phys. Rev. Lett.",
    volume = "76",
    pages = "4492",
    year = "1996"
}

@article{Li:1997rc,
    author = "Li, Bao-An and Ko, C. M. and Ren, Zhong-zhou",
    title = "{Equation of state of asymmetric nuclear matter and collisions of neutron rich nuclei}",
    eprint = "nucl-th/9701048",
    archivePrefix = "arXiv",
    doi = "10.1103/PhysRevLett.78.1644",
    journal = "Phys. Rev. Lett.",
    volume = "78",
    pages = "1644",
    year = "1997"
}

@article{Li:2003ts,
    author = "Li, Bao-An and Das, Champak B. and Das Gupta, Subal and Gale, Charles",
    title = "{Effects of momentum dependent symmetry potential on heavy ion collisions induced by neutron rich nuclei}",
    eprint = "nucl-th/0312054",
    archivePrefix = "arXiv",
    doi = "10.1016/j.nuclphysa.2004.02.016",
    journal = "Nucl. Phys. A",
    volume = "735",
    pages = "563--584",
    year = "2004"
}

@article{Hen:2014yfa,
    author = "Hen, Or and Li, Bao-An and Guo, Wen-Jun and Weinstein, L. B. and Piasetzky, Eli",
    title = "{Symmetry Energy of Nucleonic Matter With Tensor Correlations}",
    eprint = "1408.0772",
    archivePrefix = "arXiv",
    primaryClass = "nucl-ex",
    doi = "10.1103/PhysRevC.91.025803",
    journal = "Phys. Rev. C",
    volume = "91",
    number = "2",
    pages = "025803",
    year = "2015"
}

@article{PhysRevC.79.064308,
  title = {Depletion of the nuclear Fermi sea},
  author = {Rios, Arnau and Polls, Artur and Dickhoff, W. H.},
  journal = {Phys. Rev. C},
  volume = {79},
  issue = {6},
  pages = {064308},
  numpages = {17},
  year = {2009},
  month = {Jun},
  publisher = {American Physical Society},
  doi = {10.1103/PhysRevC.79.064308},
  url = {https://link.aps.org/doi/10.1103/PhysRevC.79.064308}
}

@article{PhysRevC.87.014314,
  title = {Three-body force effect on nucleon momentum distributions in asymmetric nuclear matter within the framework of the extended Brueckner-Hartree-Fock approach},
  author = {Yin, Peng and Li, Jian-Yang and Wang, Pei and Zuo, Wei},
  journal = {Phys. Rev. C},
  volume = {87},
  issue = {1},
  pages = {014314},
  numpages = {7},
  year = {2013},
  month = {Jan},
  publisher = {American Physical Society},
  doi = {10.1103/PhysRevC.87.014314},
  url = {https://link.aps.org/doi/10.1103/PhysRevC.87.014314}
}

@article{PhysRevC.93.014619,
  title = {Symmetry energy of cold nucleonic matter within a relativistic mean field model encapsulating effects of high-momentum nucleons induced by short-range correlations},
  author = {Cai, Bao-Jun and Li, Bao-An},
  journal = {Phys. Rev. C},
  volume = {93},
  issue = {1},
  pages = {014619},
  numpages = {20},
  year = {2016},
  month = {Jan},
  publisher = {American Physical Society},
  doi = {10.1103/PhysRevC.93.014619},
  url = {https://link.aps.org/doi/10.1103/PhysRevC.93.014619}
}

@article{TMEP:2022xjg,
    author = "Wolter, Hermann and others",
    collaboration = "TMEP",
    title = "{Transport model comparison studies of intermediate-energy heavy-ion collisions}",
    eprint = "2202.06672",
    archivePrefix = "arXiv",
    primaryClass = "nucl-th",
    doi = "10.1016/j.ppnp.2022.103962",
    journal = "Prog. Part. Nucl. Phys.",
    volume = "125",
    pages = "103962",
    year = "2022"
}

@article{Gan1994298,
	author = {Gan, N. and Brinkmann, K.-T. and Caraley, A.L. and Fineman, B.J. and Kernan, W.J. and McGrath, R.L. and Danielewicz, P.},
	doi = {10.1103/PhysRevC.49.298},
	journal = {Physical Review C},
	note = {},
	number = {1},
	pages = {298 -- 303},
	publication_stage = {Final},
	source = {Scopus},
	title = {Neutron-proton bremsstrahlung from low-energy heavy-ion reactions},
	type = {Article},
	url = {https://www.scopus.com/inward/record.uri?eid=2-s2.0-0002172608&doi=10.1103%2fPhysRevC.49.298&partnerID=40&md5=34557a8eed4c19b1555841e0b43837a4},
	volume = {49},
	year = {1994}}

@article{Bertsch1984sbkd,
    author = "Bertsch, G. F. and Kruse, H. and Gupta, S. D.",
    title = "{BOLTZMANN EQUATION FOR HEAVY ION COLLISIONS}",
    doi = "10.1103/PhysRevC.33.1107",
    journal = "Phys. Rev. C",
    volume = "29",
    pages = "673--675",
    year = "1984",
    note = "[Erratum: Phys.Rev.C 33, 1107--1108 (1986)]"
}

@article{Bertsch1988rev,
    author = "Bertsch, G. F. and Das Gupta, S.",
    title = "{A Guide to microscopic models for intermediate-energy heavy ion collisions}",
    doi = "10.1016/0370-1573(88)90170-6",
    journal = "Phys. Rept.",
    volume = "160",
    pages = "189--233",
    year = "1988"
}

@article{Cassing1990rev,
    author = "Cassing, W. and Metag, V. and Mosel, U. and Niita, K.",
    title = "{Production of energetic particles in heavy ion collisions}",
    doi = "10.1016/0370-1573(90)90164-W",
    journal = "Phys. Rept.",
    volume = "188",
    pages = "363--449",
    year = "1990"
}

@article{Li1995prc2037,
    author = "Li, Bao-An and Ko, Che Ming",
    title = "{Formation of superdense hadronic matter in high-energy heavy ion collisions}",
    eprint = "nucl-th/9505016",
    archivePrefix = "arXiv",
    doi = "10.1103/PhysRevC.52.2037",
    journal = "Phys. Rev. C",
    volume = "52",
    pages = "2037--2063",
    year = "1995"
}

@article{Li2005sig,
    author = "Li, Bao-An and Chen, Lie-Wen",
    title = "{Nucleon-nucleon cross sections in neutron-rich matter and isospin transport in heavy-ion reactions at intermediate energies}",
    eprint = "nucl-th/0508024",
    archivePrefix = "arXiv",
    doi = "10.1103/PhysRevC.72.064611",
    journal = "Phys. Rev. C",
    volume = "72",
    pages = "064611",
    year = "2005"
}

@article{Nifenecker1985mr,
    author = "Nifenecker, H. and Bondorf, J. P.",
    title = "{Nuclear electromagnetic bremsstrahlung: A new tool for studying heavy-ion reactions}",
    doi = "10.1016/S0375-9474(85)80028-2",
    journal = "Nucl. Phys. A",
    volume = "442",
    pages = "478--508",
    year = "1985"
}

@article{Schafer1991cme,
    author = "Schaefer, M. and Biro, T. S. and Cassing, W. and Mosel, U. and Nifenecker, H. and Pinston, J. A.",
    title = "{Photon production in nucleon-nucleon collisions}",
    doi = "10.1007/BF01560642",
    journal = "Z. Phys. A",
    volume = "339",
    pages = "391--398",
    year = "1991"
}

@article{Wang2017fnp,
    author = "Wang, Zhi and Xu, Chang and Ren, Zhongzhou and Gao, Chao",
    title = "{Probing the high-momentum component in the nucleon momentum distribution by nucleon emission from intermediate-energy nucleus-nucleus collisions}",
    doi = "10.1103/PhysRevC.96.054603",
    journal = "Phys. Rev. C",
    volume = "96",
    number = "5",
    pages = "054603",
    year = "2017"
}

@article{Yong2011sig,
    author = "Yong, Gao-Chan and Zuo, Wei and Zhang, Xun-Chao",
    title = "{A Direct probe of the in-medium pn scattering cross section}",
    eprint = "1110.2242",
    archivePrefix = "arXiv",
    primaryClass = "nucl-th",
    doi = "10.1016/j.physletb.2011.10.020",
    journal = "Phys. Lett. B",
    volume = "705",
    pages = "240--243",
    year = "2011"
}
\end{document}